\documentclass[12pt,preprint]{aastex}

\shorttitle{Stellar Rotation in M37}
\shortauthors{Hartman et al.}

\begin{document}

\title{Deep MMT\footnote{Observations reported here were obtained at the MMT Observatory, a joint facility of the Smithsonian Institution and the University of Arizona.} Transit Survey of the Open Cluster M37 III: Stellar Rotation at 550 Myr}
\author{J.~D.~Hartman\altaffilmark{2}, B.~S.~Gaudi\altaffilmark{3}, M.~H.~Pinsonneault\altaffilmark{3}, K.~Z.~Stanek\altaffilmark{3}, M.~J.~Holman\altaffilmark{2}, B.~A.~McLeod\altaffilmark{2}, S.~Meibom\altaffilmark{2}, J.~A.~Barranco\altaffilmark{4}, and J.~S.~Kalirai\altaffilmark{5,6}}
\altaffiltext{2}{Harvard-Smithsonian Center for Astrophysics, 60 Garden St., Cambridge, MA~02138, USA; jhartman@cfa.harvard.edu, mholman@cfa.harvard.edu, bmcleod@cfa.harvard.edu, smeibom@cfa.harvard.edu}
\altaffiltext{3}{Department of Astronomy, The Ohio State University, Columbus, OH~43210, USA; gaudi@astronomy.ohio-state.edu, pinsono@astronomy.ohio-state.edu, kstanek@astronomy.ohio-state.edu}
\altaffiltext{4}{Department of Physics and Astronomy, San Francisco State University, 1600 Holloway Ave., San Francisco, CA~94132, USA; barranco@stars.sfsu.edu}
\altaffiltext{5}{University of California Observatories/Lick Observatory, University of California at Santa Cruz, Santa Cruz CA, 95060; jkalirai@ucolick.org}
\altaffiltext{6}{Hubble Fellow}

\begin{abstract}
In the course of conducting a deep ($14.5 \la r \la 23$), 20 night survey for transiting planets in the rich $\sim 550$ Myr old open cluster M37 we have measured the rotation periods of 575 stars which lie near the cluster main sequence, with masses $0.2 M_{\odot} \lesssim M \lesssim 1.3 M_{\odot}$. This is the largest sample of rotation periods for a cluster older than $500~{\rm Myr}$. Using this rich sample we investigate a number of relations between rotation period, color and the amplitude of photometric variability. Stars with $M \ga 0.8 M_{\odot}$ show a tight correlation between period and mass with heavier stars rotating more rapidly. There is a group of 4 stars with $P > 15~{\rm days}$ that fall well above this relation, which, if real, would present a significant challenge to theories of stellar angular momentum evolution. Below $0.8 M_{\odot}$ the stars continue to follow the period-mass correlation but with a broad tail of rapid rotators that expands to shorter periods with decreasing mass. We combine these results with observations of other open clusters to test the standard theory of lower-main sequence stellar angular momentum evolution. We find that the model reproduces the observations for solar mass stars, but discrepancies are apparent for stars with $0.6 \la M \la 1.0 M_{\odot}$. We also find that for late-K through early-M dwarf stars in this cluster rapid rotators tend to be bluer than slow rotators in $B-V$ but redder than slow rotators in $V-I_{C}$. This result supports the hypothesis that the significant discrepancy between the observed and predicted temperatures and radii of low-mass main sequence stars is due to stellar activity.
\end{abstract}

\keywords{open clusters and associations:individual (M37) --- stars:rotation --- stars:low-mass --- stars:spots --- stars:variables:other}

\section{Introduction}

This paper is the third in a series of papers on a deep survey for transiting planets in the open cluster M37 (NGC 2099) using the MMT. In the first paper \citep[Paper I]{Hartman.07a} we introduced the survey, described the spectroscopic and photometric observations and the data reduction, determined the fundamental parameters (age, metallicity, distance and reddening) of the cluster, and obtained its mass and luminosity functions and radial density profile. In the second paper \citep[Paper II]{Hartman.07b} we identified 1445 variable stars in the field of the cluster, of which 99$\%$ were new discoveries. We found that $\ga 500$ of these variables lie near the cluster main sequence on photometric color-magnitude diagrams (CMDs) and show a correlation between period and magnitude. In this paper we investigate this population of variables arguing that the photometric variations are due to spotted star rotation. We use these variables to study the rotation of F-M main sequence stars at the age of the cluster ($550 \pm 30$ Myr, Paper I; note that for the rest of this paper we adopt this age which was derived by comparison to models with convective overshooting).

Rotation plays an important role in the life of a star. Empirically there is a strong relation between the rotation rate and the activity and age of low-mass stars \citep{Skumanich.72}, with older stars being slower rotators and less active than younger stars. Rotation affects the structure of a star \citep[see][]{Sills.00} and its evolution may have important consequences for mixing at the base of a star's convection zone \citep[see the review by][]{Pinsonneault.97}. Rotation also affects the spectral energy distribution of low-mass stars \citep{Stauffer.03}. It has even been suggested that rotation may be used as a tool to determine the ages of field stars older than a few hundred million years, and that ages determined in this fashion may be significantly more accurate than ages determined using other methods \citep{Barnes.07}. However, the usefulness of these ``gyrochronology'' ages is predicated on an accurate understanding and calibration of the age-rotation rate relation. As we discuss next, there are relatively few constraints on this relation for ages $\ga 200~{\rm Myr}$.

The surface rotation period of a star can be measured directly from photometric variations if it has substantial surface brightness inhomogeneities, from variations in the strength of spectroscopic features such as the cores of the Ca II H and K lines, or it can be constrained by measuring the projected equatorial rotation velocity $v \sin i$ from the Doppler broadening of spectral absorption lines. The latter method suffers from the inclination axis ambiguity and thus requires a substantial ensemble of stars, as well as an assumption on the inclination axis distribution, to determine the underlying velocity, and hence rotation period, distribution. While it is more desirable for the study of stellar rotation to directly measure the rotation period of a star than $v \sin i$, a consequence of the activity-age relation is that older stars show lower amplitude photometric variations. For example, at $100~{\rm Myr}$ a typical solar-like star will have a photometric amplitude of $\sim 7\%$ in $r$, at $500~{\rm Myr}$ its photometric amplitude will have fallen to $\sim 1\%$, and at $2~{\rm Gyr}$ the typical amplitude is only $\sim 0.1\%$. Moreover, longer rotation periods are more difficult to measure as they require observations spanning a longer baseline. As a result it becomes increasingly difficult to directly measure the rotation periods of stars at greater ages. Our understanding of the angular momentum evolution of stars older than a few hundred million years has been gleaned primarily from $v \sin i$ measurements (See for example the review by \citet{Bouvier.97}).

Most photometric studies of stellar rotation have focused on young ($< 100$ Myr) open clusters whose stars show relatively large amplitude photometric variations (See \citet{Stassun.03} for a review). These studies have given great insight into the evolution of angular momentum for pre-main sequence stars (PMS). As a star contracts on to the main sequence the expectation is that it will spin up. Observations of the youngest clusters, however, have revealed that not all PMS achieve these short rotation periods (See for example \citet{Herbst.02} for observations of the $\sim 1$ Myr Orion Nebula Cluster, \citet{Cieza.06} for observations of $\sim 2-3$ Myr IC 348, \citet{Lamm.05} for observations of $\sim 2-4$ Myr NGC 2264, \citet{Irwin.08b} for observations of $\sim 5$ Myr NGC 2362, and \citet{Irwin.08a} for observations of $\sim 40$ Myr NGC 2547). Possible explanations for the presence of slow rotators in these clusters include magnetic locking of the star's rotation to the inner accretion disk \citep{Konigl.91,Shu.94} or the presence of an accretion-driven wind which carries angular momentum away from the star \citep{Shu.00}.

Photometric determinations of rotation periods have been obtained for stars in a number of clusters in the age range $50-200$ Myr; at these ages solar-like stars have settled on to the main sequence. Clusters that have been studied include IC 2391 \citep[$\sim 50~{\rm Myr}$;][]{Patten.96}, IC 2602 \citep[$\sim 50~{\rm Myr}$;][]{Barnes.99}, $\alpha$ Persei \citep[$\sim 80~{\rm Myr}$;][]{Stauffer.85,Stauffer.89,Prosser.91a,Prosser.93a,Prosser.93b,ODell.93,ODell.94,Prosser.95,ODell.96,Allain.96,Martin.97,Prosser.98a,Prosser.98b,Barnes.98a}, the Pleiades \citep[$\sim 125~{\rm Myr}$;][]{Magnitskii.87,Stauffer.87a,VanLeeuwen.87,Prosser.93a,Prosser.93b,Prosser.95,Krishnamurthi.98,Terndrup.99,Scholz.04}, NGC 2516 \citep[$\sim 150~{\rm Myr}$;][]{Irwin.07}, and M34 \citep[$\sim 200~{\rm Myr}$;][]{Barnes.03,Irwin.06}. 

The data for older clusters is scarcer. Periods have been determined for 87 stars in NGC 3532 \citep[$\sim 300~{\rm Myr}$;][]{Barnes.98b}, 4 stars in Coma \citep[$\sim 600~{\rm Myr}$;][]{Radick.90}, 35 stars in the Hyades \citep[$\sim 625~{\rm Myr}$;][]{Radick.87,Radick.95,Prosser.95}, and 5 stars in Praesepe \citep[$\sim 625~{\rm Myr}$;][]{Scholz.07}.

The observations discussed above have clearly demonstrated that once a low-mass star (G and later) reaches the main sequence it begins to lose angular momentum; this is understood to be the result of a magnetized stellar wind \citep{Webber.67}. By comparing the rotation velocities of stars in the Pleiades and the Hyades with the Sun, \citet{Skumanich.72} found the scaling relation $v \propto t^{-1/2}$, where $v$ is the average equatorial velocity and $t$ is the age. This can be explained by a surface angular momentum loss rate that is proportional to $\omega^3$ \citep{Kawaler.88} and naturally leads to a convergence in the rotation rates of stars at a given mass as seen by \citet{Radick.87} for stars in the Hyades. The presence of solar-mass rapid rotators in the Pleiades \citep{VanLeeuwen.82} appears to contradict the \citet{Skumanich.72} law and suggests that the angular momentum loss rate is saturated above some critical rotation rate $\omega_{crit}$ so that it scales as $\omega_{crit}^2\omega$ for $\omega > \omega_{crit}$ \citep{Kawaler.88}. The critical rotation rate must be mass dependent to explain the spread in rotation rates for lower mass stars \citep{Krishnamurthi.97}. 

The rotation history of a star will depend not only on the rate of angular momentum loss, but also on the efficiency of internal angular momentum transport. Models in which the core and envelope of a star are decoupled lead to a rapid spin-down of the envelope \citep{Soderblom.93}. In these models once the core and envelope become re-coupled the spin-down is much less dramatic. Models in which the internal angular momentum transport is calculated by hydrodynamic mechanisms, and in which some degree of core-envelope decoupling is permitted have been produced (e.g. \citet{Sills.00} show that models that incorporate differential rotation with depth in stars are needed to reproduce the angular momentum evolution of systems younger than the Pleiades). Alternatively, models in which the star is assumed to rotate as a solid body have also been developed (e.g. \citet{Bouvier.97b}).

This paper presents rotation periods for 575 stars in the open cluster M37 (NGC 2099). In Paper I we found that the cluster has an age of $550 \pm 30$ Myr, a reddening of $E(B-V) = 0.227 \pm 0.038$, a distance modulus of $(m-M)_{V} = 11.57 \pm 0.13$ and a metallicity of $[M/H] = +0.045 \pm 0.044$ which are in good agreement with previous measurements. The age of M37 is thus comparable to that of the Hyades \citep[625 Myr with overshooting;][]{Perryman.98} which at present is the oldest cluster for which a significant number of stellar rotation periods have been measured. M37 is, however, substantially richer than the Hyades and thus has the potential to provide a much larger data-set of rotation periods for older stars. 

As we were preparing to submit this manuscript, we became aware of a similar, independent study by \citet{Messina.08}, which presents rotation periods for 122 stars in this cluster. Our survey goes more than 2 magnitudes deeper than \citet{Messina.08} with more than an order of magnitude more observations from more than twice as many nights. As a result we are able to study the rotation evolution for late K and early M dwarfs as well as the late F, G and early K dwarfs studied by \citet{Messina.08}. On the other hand, the \citet{Messina.08} survey has measured periods for early F stars that are saturated in our survey. We compare the periods measured by both surveys in \S 3.6.

In the next section we provide a brief summary of the observations and data reduction. In \S 3 we compile the catalog of candidate cluster members with measured rotation periods. In \S 4 we fit analytic models to the observed period-color sequence in M37. In \S 5 we compare the photometric observations to spectroscopic $v \sin i$ measurements for a number of these stars. In \S 6 we study the amplitude distribution as a function of period and color. In \S 7 we study the Blue K dwarf phenomenon in M37. In \S 8 we compare these observations to theories of stellar angular momentum evolution. We discuss the results in \S 9.

\section{Summary of Observations}

The observations and data reduction procedure were described in detail in Papers I and II, we provide a brief overview here. The observations consist of both $gri$ photometry for $\sim 16000$ stars, and $r$ time-series photometry for $\sim 23000$ stars obtained with the Megacam mosaic imager \citep{McLeod.00} on the 6.5 m MMT. We also obtained high-resolution spectroscopy of 127 stars using the Hectochelle multi-fiber, high-dispersion spectrograph \citep{Szentgyorgyi.98} on the MMT.

The primary time-series photometric observations were done using the $r$ filter and consist of $\sim 4000$ high quality images obtained over twenty four nights (including eight half nights) between December 21, 2005 and January 21, 2006. We obtained light curves for stars with $14.5 \la r \la 23$ using a reduction pipeline based on the image subtraction technique and software due to \citet{Alard.98} and \citet{Alard.00}. We apply two cleaning routines to the data: clipping outlier points and removing individual bad images. We do not decorrelate against other systematic variations since doing so tends to distort the light curves of large amplitude variables.

The spectra were obtained on four separate nights between February 23, 2007 and March 12, 2007 and were used to measure $T_{eff}$, $[Fe/H]$, $v\sin i$ and the radial velocity (RV) via cross-correlation against a grid of model stellar spectra computed using ATLAS 9 and SYNTHE \citep{Kurucz.93}. The classification procedure was developed by Meibom et al. (2008, in preparation), and made use of the \emph{xcsao} routine in the {\scshape Iraf}\footnote{{\scshape Iraf} is distributed by the National Optical Astronomy Observatories, which is operated by the Association of Universities for Research in Astronomy, Inc., under agreement with the National Science Foundation.} \emph{rvsao} package \citep{Kurtz.98} to perform the cross-correlation. In performing the cross-correlation we found that it was necessary to fix $\log(g) = 4.5$, however given that very few of the field stars in our sample are likely to be giants or sub-giants fixing $\log(g)$ should not substantially bias the resulting parameters. We measured the parameters separately for each of the four nights choosing the $v\sin i$ and $T_{eff}$ values that maximize the cross-correlation peak-height value at three $[Fe/H]$ grid points ($[Fe/H] = -0.5$, $[Fe/H] = 0.0$, and $[Fe/H] = +0.5$). For each $[Fe/H]$ grid point we determined the average $v\sin i$ and $T_{eff}$ values over the four nights together with uncertainties on the values for each night estimated using the standard deviation of the measurements. We then fit a quadratic relation between $[Fe/H]$ and the average peak-height values to estimate the value of $[Fe/H]$ that maximizes the cross-correlation and then determined $v\sin i$ and $T_{eff}$ by fitting a quadratic relation between each of these parameters and $[Fe/H]$. The final errors on each parameter are the standard errors from the fit. For a given star the error on $v\sin i$ can be substantially larger than the average value, particularly for fainter targets, when the value of $[Fe/H]$ is not strongly constrained or when $v\sin i$ is small, in these cases the error should be taken as an upper limit on the value. To determine the $RV$ for each star we used \emph{rvsao} to cross-correlate the spectra on each night against the best matching template spectrum from the full grid.

As described in Paper I we also take $BV$ photometry for stars in the field of this cluster from \citet{Kalirai.01}, $K_{S}$ photometry from 2MASS \citep{Skrutskie.06} and we transform our $ri$ photometry to $I_{C}$ using a transformation based on the $I_{C}$ photometry from \citet{Nilakshi.02}.

\section{Catalog of Candidate Cluster Members with Measured Rotation Periods}
\subsection{Selection of Rotational Variables}\label{sec:rotsel}

\begin{figure}[p]
\plotone{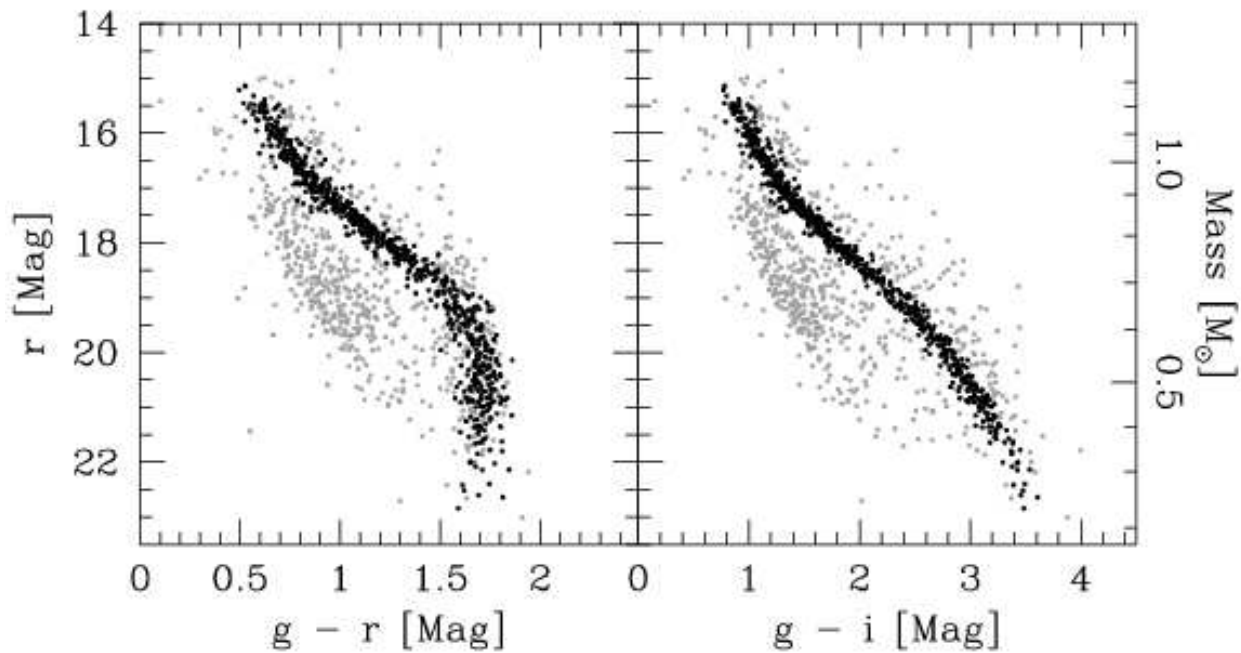}
\caption{$g-r$ vs $r$ (left) and $g-i$ vs $r$ (right) CMDs showing the selection of variable stars that are candidate cluster members; only variable stars are included on this plot. Dark points show the candidate cluster member variables, light points show field variables. Variables classified as pulsators or eclipsing binaries (see Paper II) are not included on this plot. The vertical scale on the right side of the plot shows the masses of cluster members as a function of $r$ magnitude.}
\label{fig:VarSelect}
\end{figure}

In Paper II we selected 1445 periodic variable stars using the Lomb-Scargle \citep[L-S;][]{Lomb.76, Scargle.82, Press.89, Press.92}, phase-binning analysis of variance \citep[AoV;][]{Schwarzenberg-Czerny.89, Devor.05} and box-fitting least squares \citep[BLS;][]{Kovacs.02} algorithms. From this catalog we will now select a population of probable cluster member, rotational variables. To do this we follow the procedure for selecting candidate photometric cluster members described in Paper I. For each star we determine the $g,r,i,B,V$ point within the fiducial cluster main sequences, generated by eye, that has a minimum $\chi^{2}$ deviation from the observed $g,r,i,B,V$ values for the star. We then select as candidate cluster members stars with $\chi^{2} < 150$ in $g,r,i$ and $\chi^{2} < 250$ in $B,V$ for $B-V < 1.38$ and $V < 20$ or $\chi^{2} < 150$ in $B,V$ for other stars. Figure~\ref{fig:VarSelect} shows the selected variables on $g-r$ and $g-i$ CMDs. After rejecting variables that were classified in Paper II as eclipsing binaries or short period pulsators and rejecting variables without period determinations we are left with 575 variables that are candidate cluster members. A catalog of these 575 variables is given in tables~\ref{tab:M37_rotation1}-\ref{tab:M37_rotation3}, we show only the first ten rows of each table for illustration, the full tables are available in the online edition of the journal. Note that only stars with spectroscopy are included in table~\ref{tab:M37_rotation3}.

\subsection{Revised Periods}

\begin{figure}[p]
\plotone{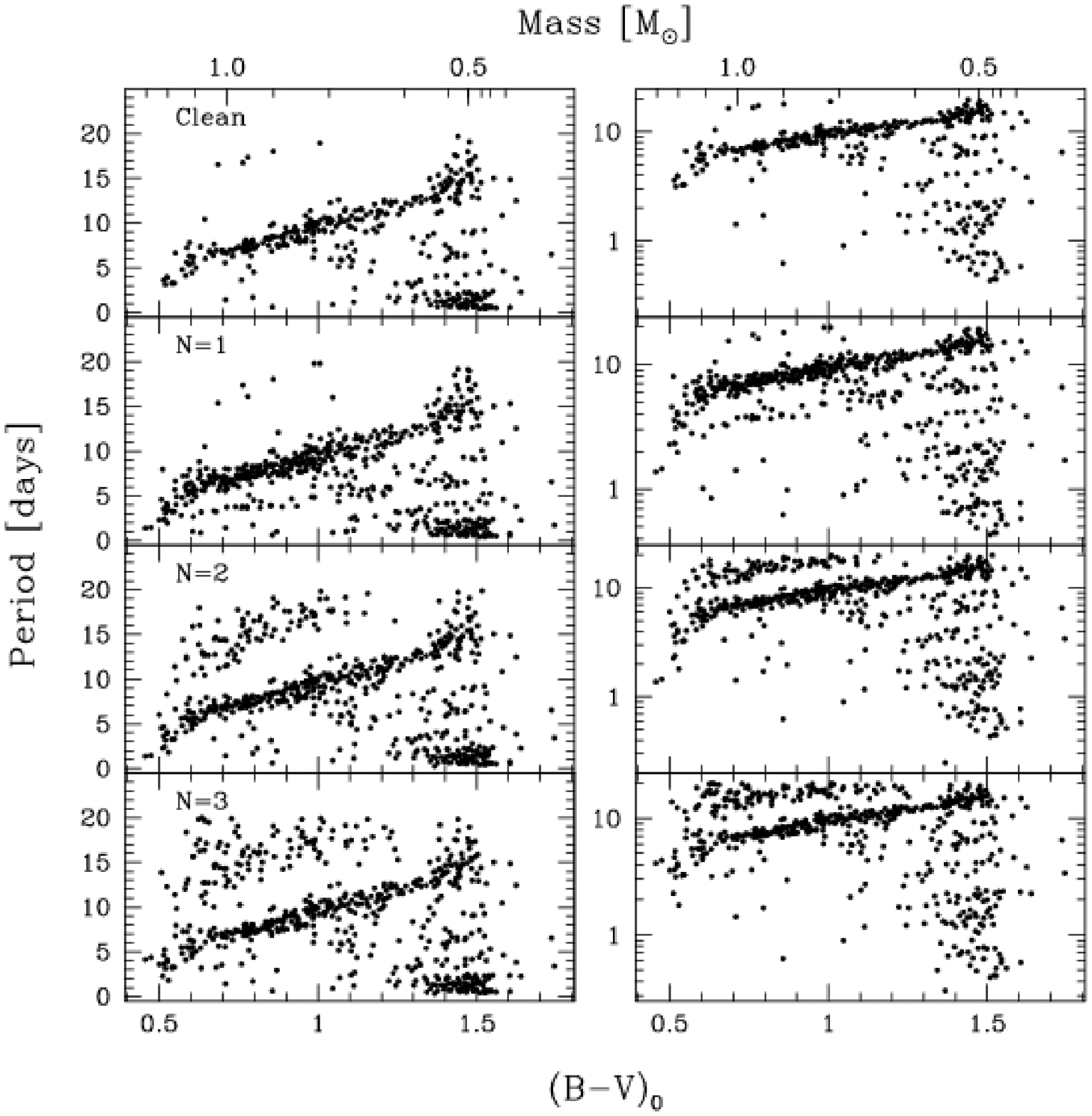}
\caption{Period versus $(B-V)_{0}$ color (bottom axis) and mass (top axis) for variable stars that are candidate cluster members. The periods are determined with the multi-harmonic AoV method, which is equivalent to fitting equation~\ref{eqn:fourier} to the light curves. We show the results for $N = 1, 2$ and $3$. In the top panel we show a clean sample of stars for which the $N = 1, 2$ and $3$ periods do not differ from each other by more than $10\%$. In the left panels the period is plotted on a linear scale, in the right panels it is plotted on a logarithmic scale. We adopt the $N=2$ period for the clean sample of stars.}
\label{fig:BVPeriod_multiharm}
\end{figure}

\begin{figure}[p]
\plotone{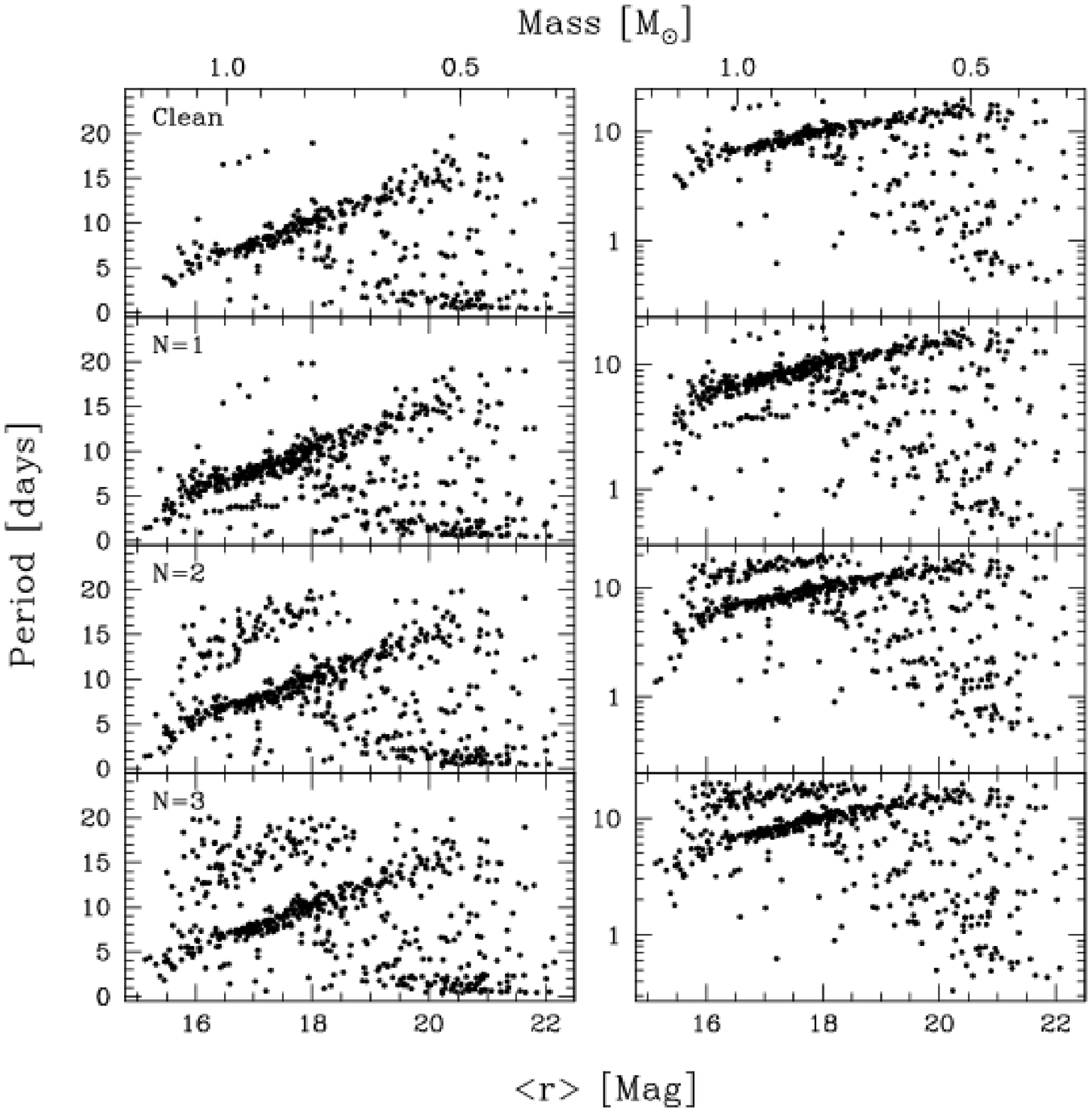}
\caption{Same as figure~\ref{fig:BVPeriod_multiharm}, here we plot period against $<r>$, the average $r$ magnitude of each light curve.}
\label{fig:AverPeriod_multiharm}
\end{figure}

In Paper II we calculated the periods of the variable stars using the L-S, phase-binning AoV, and BLS algorithms. We then selected, by eye, the most likely period for each star from the values returned by the three algorithms. Here we recalculate the periods using the multi-harmonic AoV algorithm due to \citet{Schwarzenberg-Czerny.96}. This method is equivalent to fitting to each light curve a harmonic series of the form:

\begin{equation}
\tilde{r}(t) = a_{0} + \sum_{i=1}^{N}a_{i}\cos(2\pi i t / P + \phi_{i})
\label{eqn:fourier}
\end{equation}
where $P$ is the period, $a_{i}$ are the amplitudes, and $\phi_{i}$ are the phases. This method is more general than the L-S algorithm which is equivalent to fitting a $N=1$ series without a floating mean, and may give a more accurate period determination for light curves that have multiple minima in a single cycle. We calculate the period for each light curve for $N = 1, 2$ and $3$. Figure~\ref{fig:BVPeriod_multiharm} shows the period-$(B-V)_{0}$ relation for each value of $N$ while figure~\ref{fig:AverPeriod_multiharm} shows the period-$<\!r\!>$ relation. Here $<\!r\!>$ is the average $r$ magnitude of the light curve. The magnitudes and colors have been converted to masses using the mass-$r$ and mass-$(B-V)_{0}$ relations for this cluster that were determined in Paper I.

As seen in figures~\ref{fig:BVPeriod_multiharm} and \ref{fig:AverPeriod_multiharm}, there is a clear period-stellar mass sequence running from $P \sim 3~{\rm days}$, $M \sim 1.2 M_{\odot}$ to $P \sim 17~{\rm days}$, $M \sim 0.5 M_{\odot}$. There also appears to be a significant number of stars with periods falling below the sequence for $M < 0.8 M_{\odot}$ and a cluster of stars with $15~{\rm days} < P < 20~{\rm days}$ and $0.7 M_{\odot} < M < 1.1 M_{\odot}$. For $N = 1$ there appears to be a second sequence with periods that are half the main-band values, while $N = 2$ and $N = 3$ yield a sequence with periods that are twice the main-band values. Light curves that fall in the long period sequence for $N = 2$ and $N = 3$ show multiple minima in a cycle when phased at the long period. Changing the number of harmonics in the fit can select different harmonics of the true period. Light curves with multiple, unequal, minima in a cycle or light curves that are not perfectly periodic (e.g. due to spot evolution, or uncorrected systematic errors in the photometry) are particularly susceptible to choosing an incorrect harmonic of the true period. For the remainder of the paper it is important to have a sample of stars with unambiguous periods. We therefore select a clean sample of 372 stars for which periods determined with $N=1,2$ and $3$ do not differ by more than $10\%$. For this sample we adopt the $N=2$ periods. As seen in figure~\ref{fig:BVPeriod_multiharm} and \ref{fig:AverPeriod_multiharm}, the half-period harmonic sequence is removed from the clean sample. We note that a handful of long-period variables remain in the clean sample, we discuss these further in \S 9. Tables~\ref{tab:M37_rotation1}~and~\ref{tab:M37_rotation2} include the full sample of stars, the $N = 1$, $2$ and $3$ periods are provided in table~\ref{tab:M37_rotation1}.

\begin{figure}[p]
\plotone{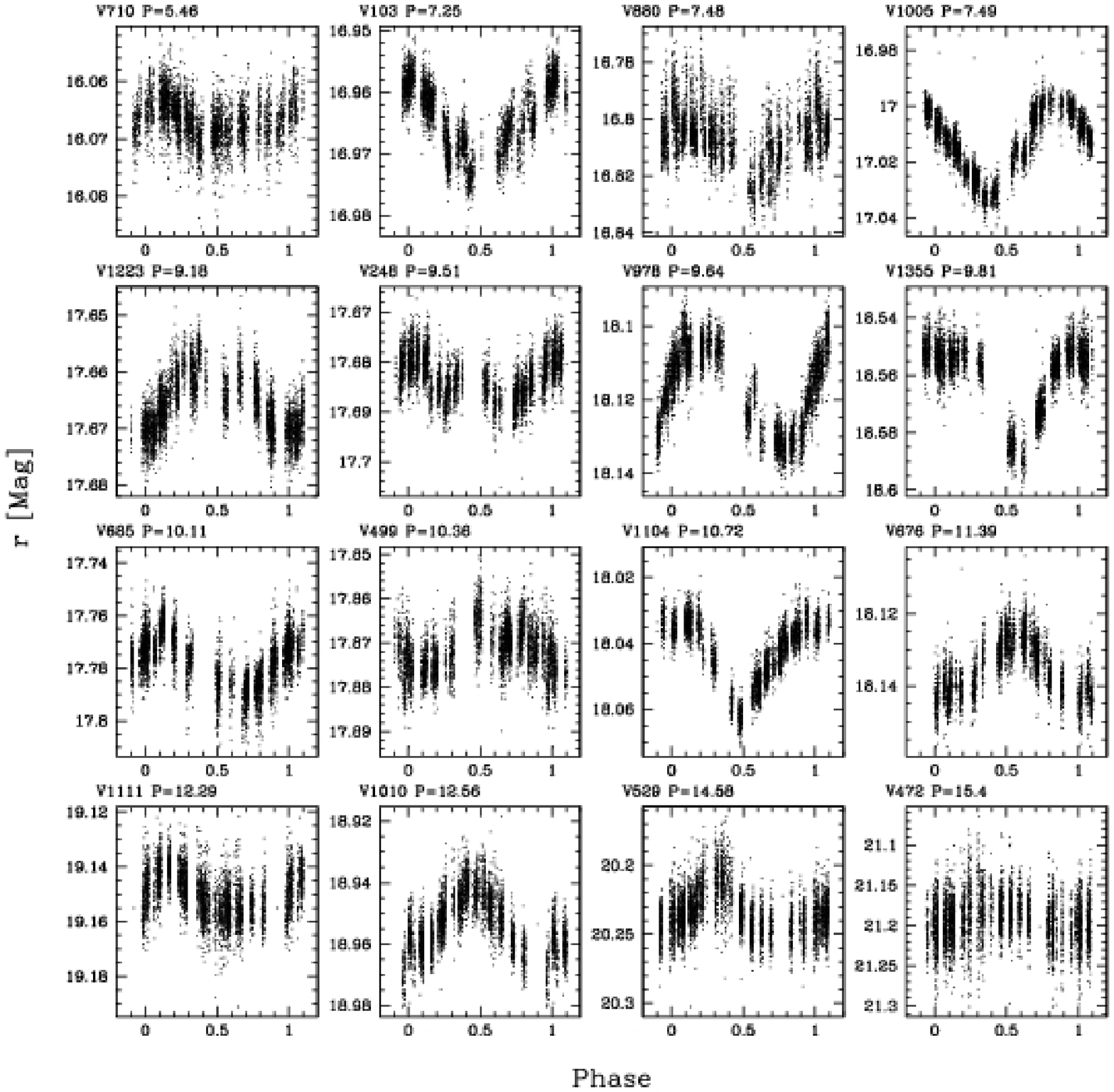}
\caption{Example light curves for 16 randomly selected variable stars that are candidate cluster members and lie along the main period-color band (\S 4). The period listed for each light curve is in days. The continuous quasi-periodic variations are consistent with spotted star rotation. The light curves are sorted by period.}
\label{fig:MainBandLC}
\end{figure}

\begin{figure}[p]
\plotone{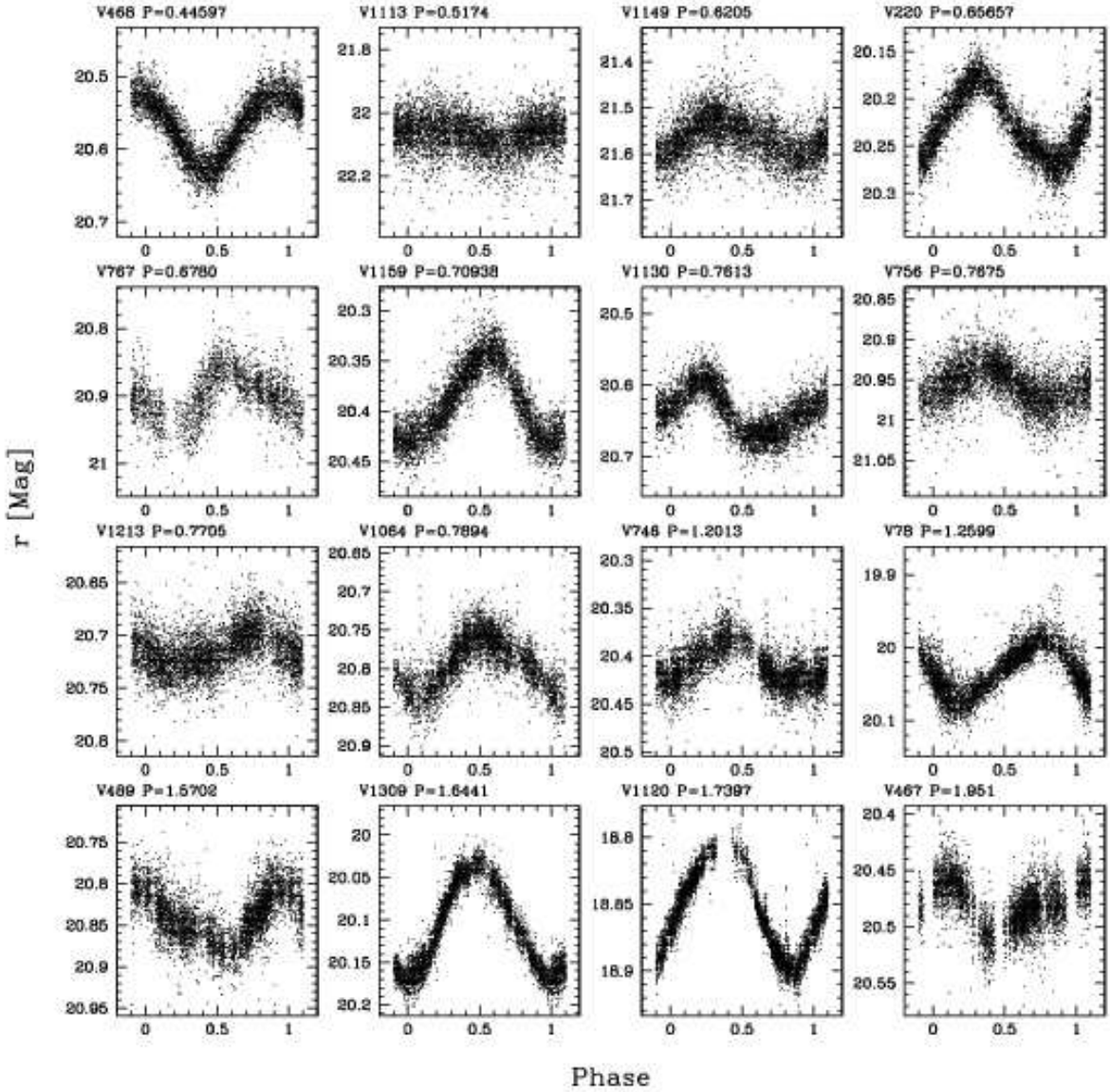}
\caption{Example light curves for 16 randomly selected candidate cluster member variable stars with $P < 2~{\rm days}$.}
\label{fig:RapidRotLC}
\end{figure}

\begin{figure}[p]
\plotone{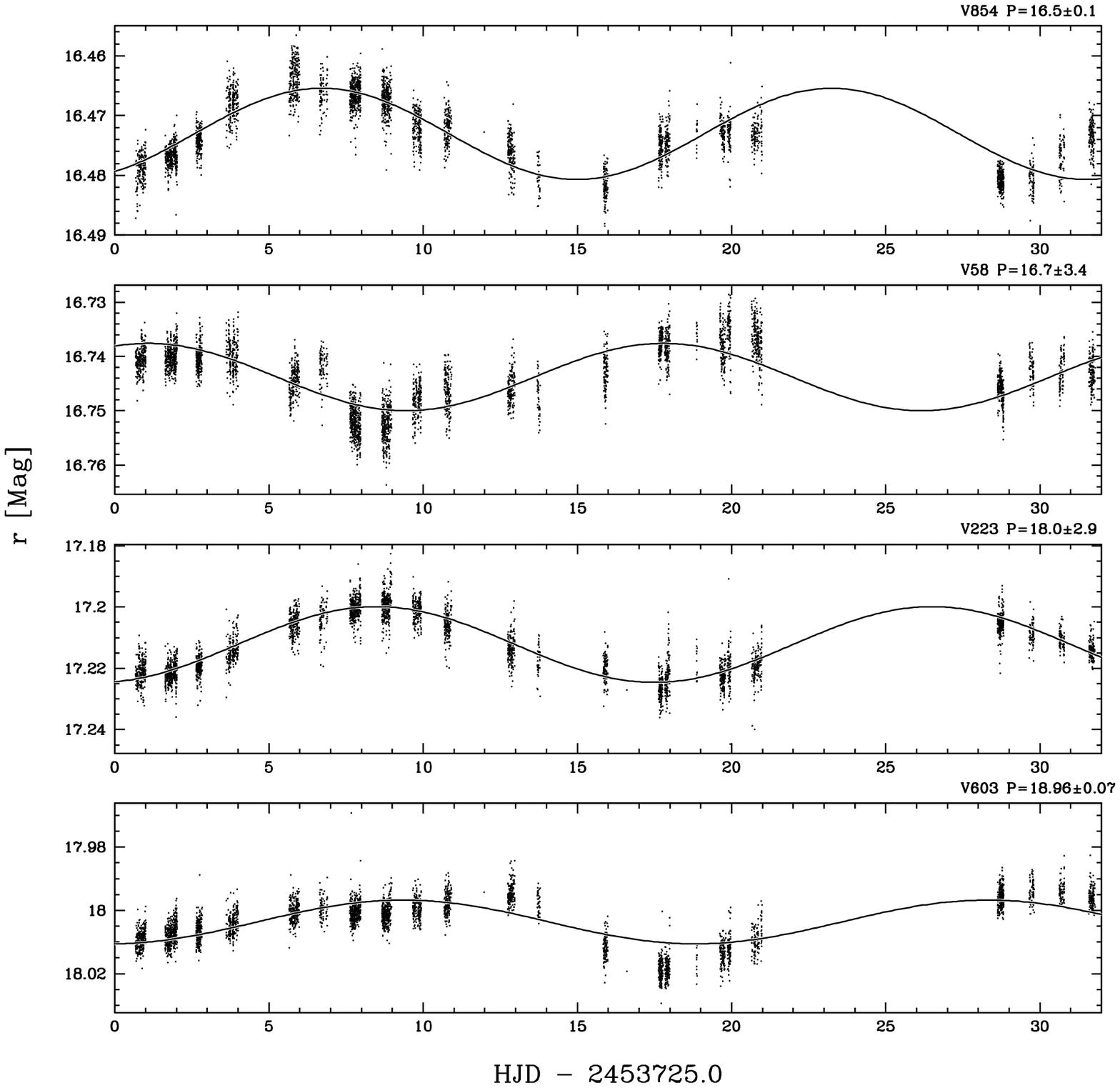}
\caption{Light curves of four variable stars that are photometrically selected candidate cluster members with long periods ($P > 15~{\rm days}$ and $(B-V)_{0} < 1.2$). The listed periods are in days and the solid line shows the best fit sinusoid to each light curve. V223 has an RV that is consistent with cluster membership; we do not have RV information for the other three stars. We exclude one long period star that is in the clean sample shown in figure~\ref{fig:BVPeriod_multiharm} but has an RV that is inconsistent with it being a member of the cluster.}
\label{fig:LongPeriodRotsLCS}
\end{figure}

In figure~\ref{fig:MainBandLC} we show phased light curves for a random sample of stars that fall along the main period-color band (see \S 4). Figure~\ref{fig:RapidRotLC} shows phased light curves for a random sample of rapid rotators ($P < 2~{\rm days}$). Finally, in figure~\ref{fig:LongPeriodRotsLCS} we show the light curves of 4 of the 5 long-period stars in the clean sample with $P > 15~{\rm days}$ and $r < 18.5~{\rm mag}$. Note that the fifth star is rejected as a non-cluster member based on its RV.

\subsection{Period and Color Uncertainties}

As we will discuss in \S 8.2, the spread in rotation periods for stars of a given mass puts a powerful constraint on theories of stellar angular momentum evolution. To determine whether or not the observed period spread is real it is important to have accurate estimates for both the period and color uncertainties. 

\subsubsection{Period Uncertainties}\label{sec:perunc}

Uncertainties in the period can be divided into two classes: errors due to aliasing or choosing the wrong harmonic of the true period and random errors. Aliasing generally yields a discrete set of peaks in the periodogram of a light curve resulting in an uncertainty in choosing the peak that corresponds to the physical period of the star. Random errors correspond to the uncertainty in the centroid of each periodogram peak, these errors are typically assumed to be Gaussian. There are at least four factors which contribute to random uncertainties in the stellar rotation period inferred from starspots. These factors include:
\begin{enumerate}
\item Noise in the photometry.
\item Inadequacies in the model used to determine the period (e.g. the light curve is periodic but not sinusoidal).
\item Spot evolution.
\item Differential rotation.
\end{enumerate}

In Paper II we conducted bootstrap simulations of the period detection for each star. This effectively determines the contribution of photometric noise to the period uncertainty. We redo these bootstrap simulations here for the multi-harmonic AoV period determinations. 

\begin{figure}[p]
\plotone{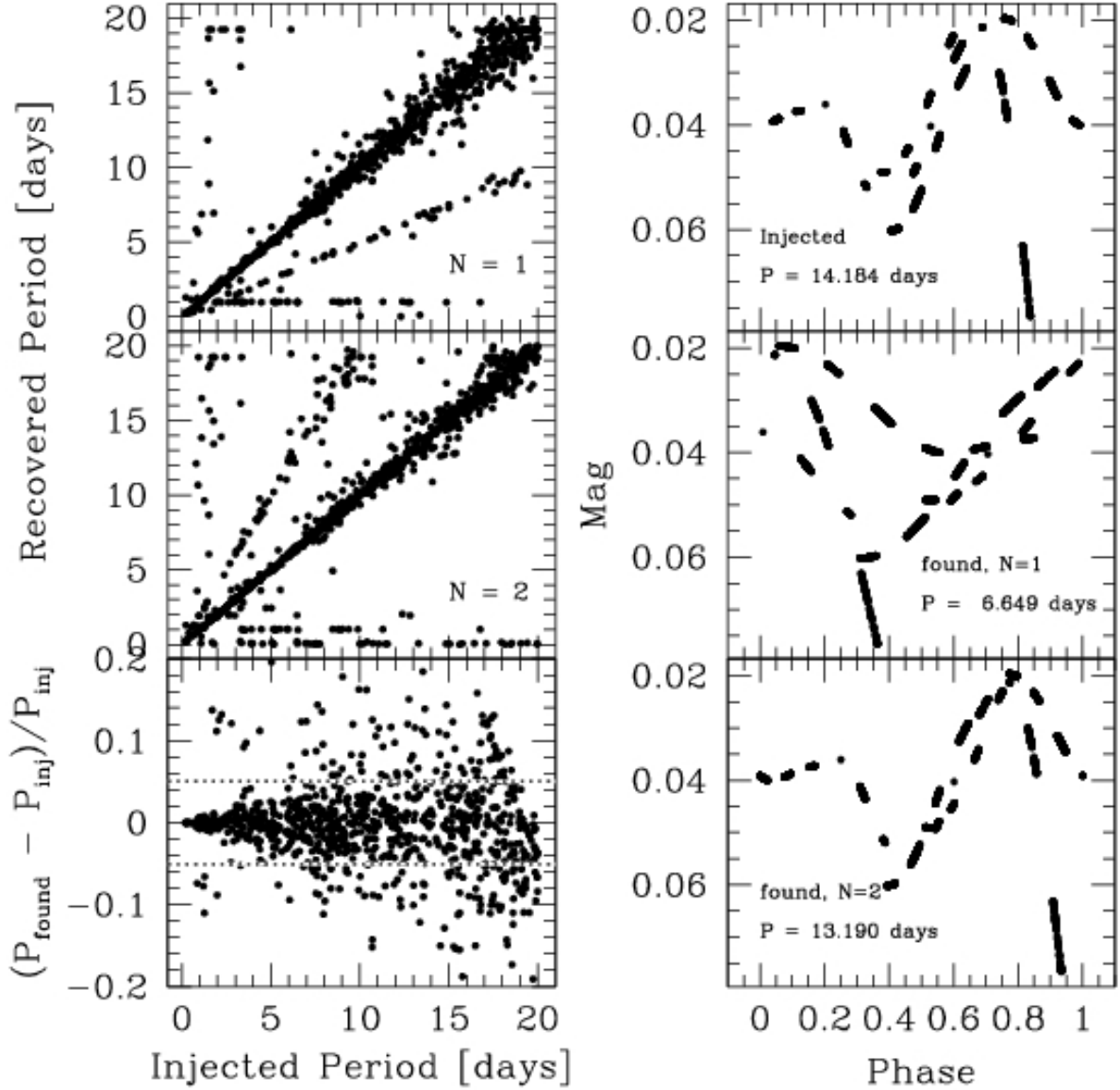}
\caption{Left: Injected versus recovered period for simulated spot evolution light curves. The periods are recovered using the multi-harmonic AoV algorithm with $N=1$, $2$ and $3$. We plot the results for $N=1$ and $N=2$ only. The bottom left panel shows the fractional difference between the recovered and injected periods for $N=2$. The dotted line shows the $RMS$. Right: A simulated light curve of a star exhibiting spot evolution is phased at the injected rotation period (top) as well as at the recovered periods for $N=1$ and $N=2$.}
\label{fig:spotsimulation}
\end{figure}

To assess the uncertainty due to inadequacies in the model and due to spot evolution we conduct Monte Carlo simulations. We simulate light curves for 1000 spotted stars using the spot model due to \citet{Dorren.87}. For each simulation we place three spots with random angular sizes between $0.05$ and $0.5$ radians, random latitudes and random longitudes on the surface of a star, assign the star a random rotation period between $0.2$ and $20~{\rm days}$, and allow the spots to vary sinusoidally in angular size. The amplitude, phase and period of the variation are chosen randomly for each spot. We limit the period for spot size variation to lie between 5 and 20 times the rotation period. We generate light curves for each simulated star using the same time sampling as our observations. We then measure the period of each simulated light curve using the multi-harmonic AoV algorithm, rejecting light curves for which $N=1,2$ and $3$ do not return the same period within $10\%$. Figure~\ref{fig:spotsimulation} shows an example of a simulated light curve and the detected period as a function of the injected period for all simulations. We find the RMS difference between the injected and recovered periods is $\sim 5\%$. We also find that the $N=2$ period has the lowest RMS deviation from the injected rotation period. Note that this estimate for the error is conservative since most of the stars that we observed do not show as much spot evolution as our models do.

The rotation periods measured for an ensemble of stars with spots located at different latitudes may exhibit scatter even if all stars of a given mass have the same equatorial rotation period. The Sun exhibits differential rotation that can be modeled as
\begin{equation}
P_{\beta} = P_{EQ}/(1 - k \sin^{2}\beta)
\end{equation}
where $P_{\beta}$ is the rotation period at latitude $\beta$ and $P_{EQ}$ is the equatorial rotation period. Observations of spots on the Sun yield $k = 0.19$ while surface radial velocity measurements give $k = 0.12$ \citep{Kitchatinov.05}. Theoretical simulations of turbulent convection predict that $k$ should decrease with increasing rotation rate \citep{Brown.04}, this has been confirmed by observations of starspots on $\kappa^{1}$ Ceti by the MOST satellite which yield $k = 0.09$ for this solar-like star rotating with $P = 8.77~{\rm days}$ \citep{Walker.07}. While sunspots are rarely seen with $|\beta| > 30^{\circ}$, there are indications that younger stars may have spots at any latitude. For example, the aforementioned observations of $\kappa^{1}$ Ceti found 7 spots over the range $10^{\circ} < |\beta| < 80^{\circ}$. Assuming spots may be uniformly distributed over the surface of a star (i.e. uniformly distributed in $\sin\beta$), a value of $k = 0.09$ will yield an RMS spread in detected periods of $\sim 3\%$. We adopt this as an estimate for the expected contribution of differential rotation to the period uncertainty for stars in M37.

\subsubsection{Color Uncertainties}

There are several effects that contribute to the uncertainty in the color of a star including uncertainties in the photometric precision, photometric variability and binarity. All of these effects can cause the measured color to differ from the photosphere color of the star (or the star with the measured rotation period in the case of a multiple star system). The net uncertainty from all of these effects can be determined empirically from the spread of the main sequence on a CMD. Following the procedure described in \S 7, we draw a fiducial main sequence by eye through the rotational variables plotted on $B-V$, $V-I_{C}$, $g-r$ and $g-i$ CMDs. We then calculate for each variable the difference between its observed color and the color along the fiducial main sequence interpolated at the $V$ magnitude of the star. The data are binned in magnitude and we calculate the RMS of the color residuals for each bin. The resulting uncertainties in $(B-V)_{0}$, $(V-I_{C})_{0}$, $g-r$ and $g-i$ as a function of absolute magnitude are listed in table~\ref{tab:coloruncertainties}. The drop in the RMS of the color residuals at $M_{V} = 11$ for $(V-I_{C})_{0}$, $g-r$ and $g-i$ is due to incompleteness in the selection of probable members at this magnitude.

\subsection{Completeness}\label{sec:completeness}

Incompleteness in the selection of variable stars can bias the observed period-color and period-amplitude distributions. To assess the completeness of our sample of rotation periods we conduct Monte Carlo simulations of the variable star selection process. We inject sinusoidal variations, and spot signals (generated using the model described in \S~\ref{sec:perunc}) into the light curves of 1081 stars that pass the cuts used to select candidate cluster members discussed in \S~\ref{sec:rotsel} but were not selected as variable stars. For each star we conduct 1000 simulations. For the sinusoid models we inject signals with random phases, periods uniformly distributed in logarithm between $0.1$ and $20.0~{\rm days}$ and semi-amplitudes uniformly distributed in logarithm between $1~{\rm mmag}$ and $0.1~{\rm mag}$. We attempt to recover the injected signals using the L-S algorithm in combination with the multi-harmonic AoV algorithm. An injected signal is considered to be recovered if it does not have a period falling within one of the rejected period bins discussed in Paper II, has a formal L-S false alarm probability logarithm that is less than $-150$ and if the N=1, 2 and 3 periods returned by AoV agree with one another to within $10\%$. To make the problem computationally feasible we use a lower period resolution in the L-S algorithm than what we used to select the variables in Paper II (we sample at $0.1$ times the Nyquist frequency rather than $0.005$ times the Nyquist frequency).

\begin{figure}[p]
\plotone{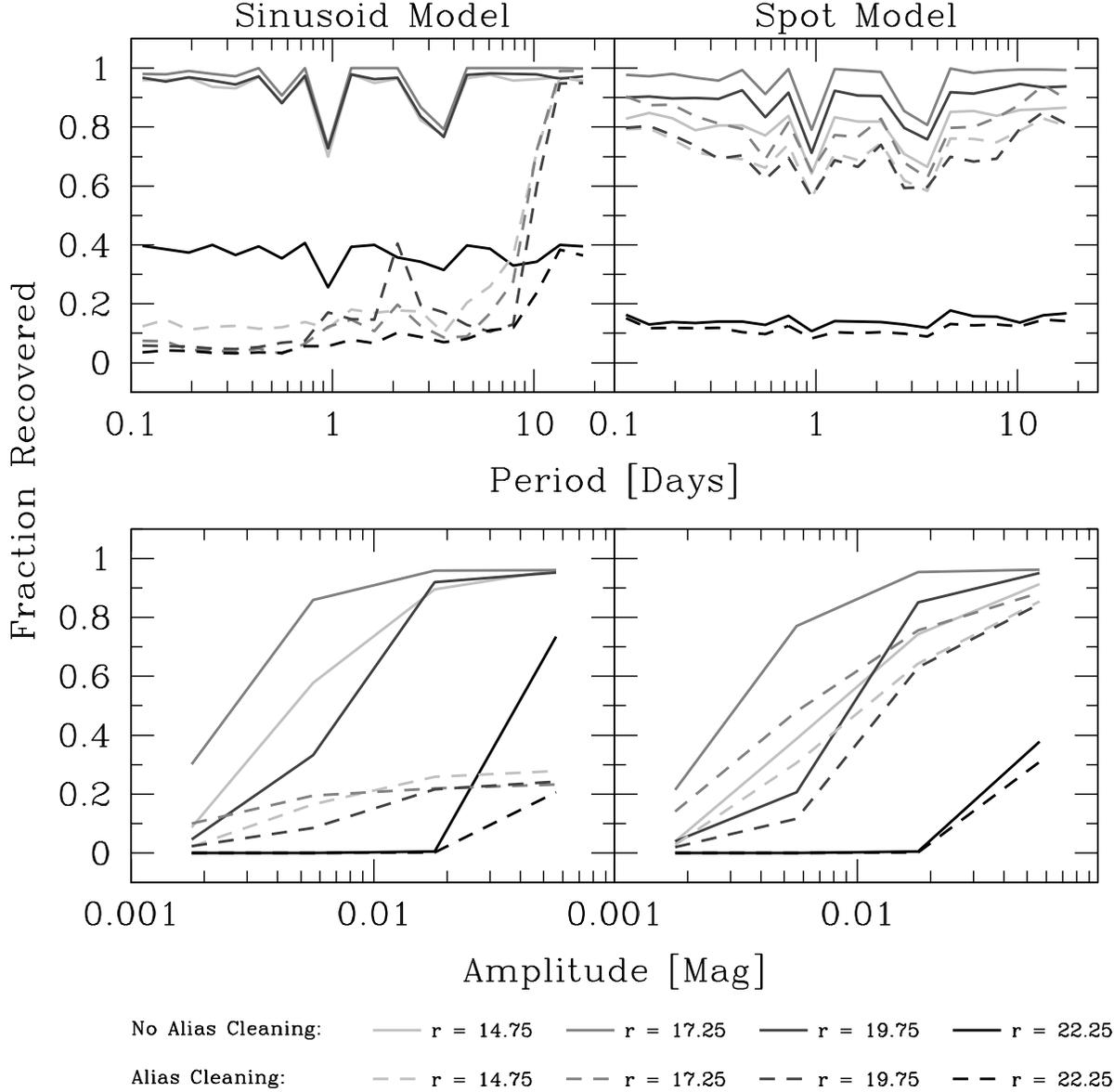}
\caption{The recovery fraction for injected sinusoid (left) and spot (right) models as a function of period (top) and amplitude (bottom). We show the results for several magnitudes with and without applying the multi-harmonic AoV selection. For the top plot we include all simulations with semi-amplitudes greater than $0.01~{\rm mag}$.}
\label{fig:RotationRecovery}
\end{figure}

\begin{figure}[p]
\plotone{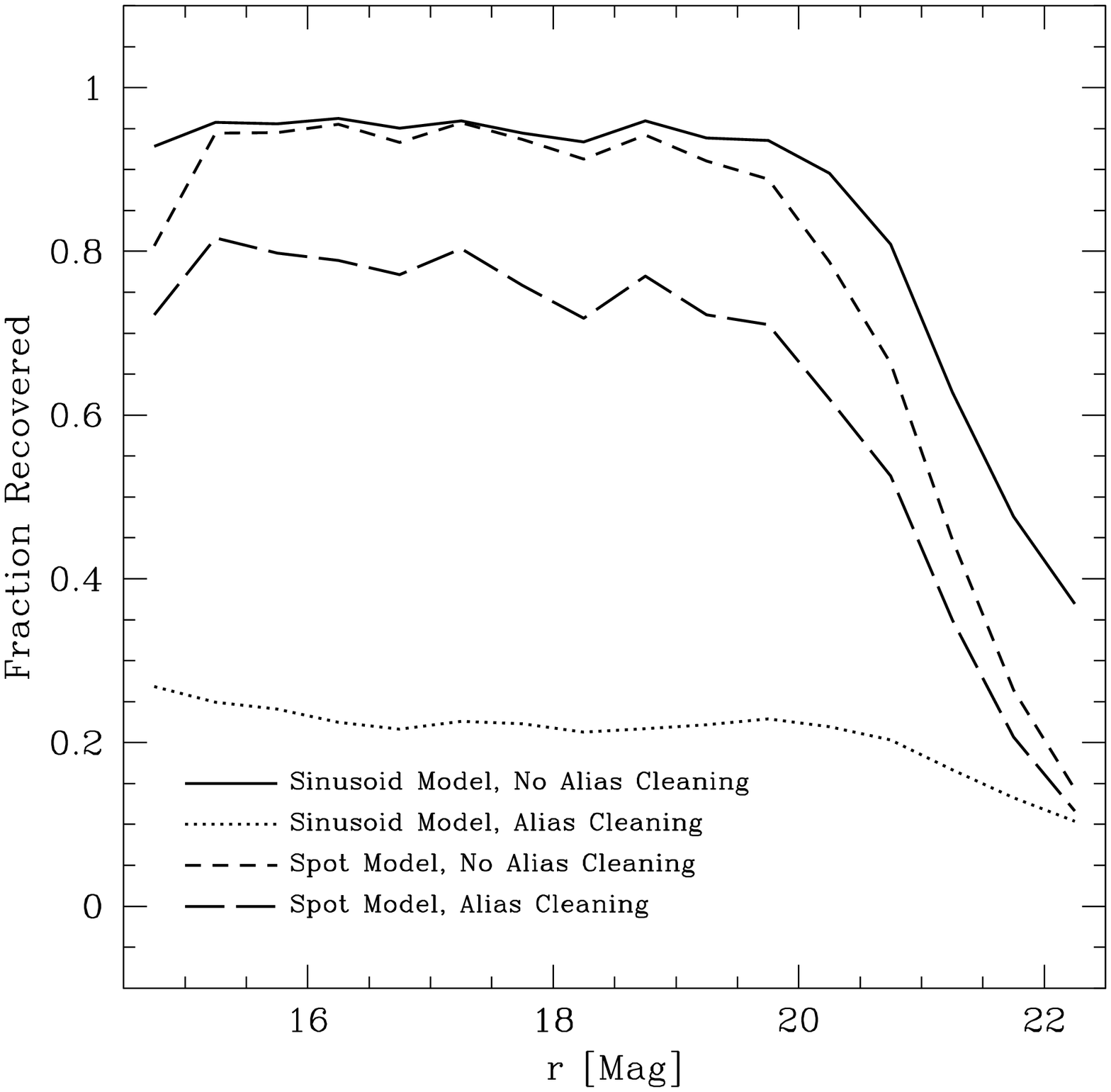}
\caption{The recovery fraction as a function of magnitude for simulations that have periods uniformly distributed in logarithm between $0.1$ and $20~{\rm days}$, and semi-amplitudes uniformly distributed in logarithm between $0.01$ and $0.1~{\rm mag}$.}
\label{fig:RotationRecovery_mag}
\end{figure}

In figure~\ref{fig:RotationRecovery} we show the recovery fraction as a function of period and of amplitude for the sinusoid injections and the spot model injections while figure~\ref{fig:RotationRecovery_mag} shows the recovery fraction as a function of magnitude. When we don't apply the multi-harmonic AoV selection, the recovery fraction is more or less insensitive to period between $0.1$ and $20$ days for semi-amplitudes larger than $0.01$ mag. The recovery fraction for stars brighter than $r \sim 20$ is close to $100\%$ for the sine-curve model and is between $80-100\%$ for the spot model. The recovery fraction drops at a few periods ($\sim 1~{\rm day}$ and $2-4~{\rm days}$) due to the period rejection that we perform in selecting the variables. For the sinusoid models applying the multi-harmonic AoV selection reduces the recovery fraction dramatically to $10-20\%$. This is due to the degeneracy between harmonic number and period (i.e. doubling the period and setting the amplitude of the first harmonic to one and all other harmonics to zero yields the same signal as using the correct period and setting the amplitude of the fundamental to one and all harmonics to zero). Because we limit the period search to 20 days, periods longer than 10 days cannot be recovered at twice the period or three times the period, and periods longer than 7 days cannot be recovered at three times the period. As a result the recovery fraction for long periods using the multi-harmonic AoV selection is close to the recovery fraction when the AoV selection is not used. For more realistic spot models the degeneracy between choosing the fundamental mode with the true period or either of the harmonics with a longer period is broken, so applying the multi-harmonic AoV selection does not have as significant an effect on the recovery fraction. In this case the recovery fraction is reduced by $\sim 20\%$ independent of period. Based on the results for the spot model signals we conclude that the rotation period-color diagram may be biased toward brighter stars, but at fixed magnitude, it is not strongly biased in period due to incompleteness. 

\subsection{Field Contamination}

\begin{figure}[p]
\plotone{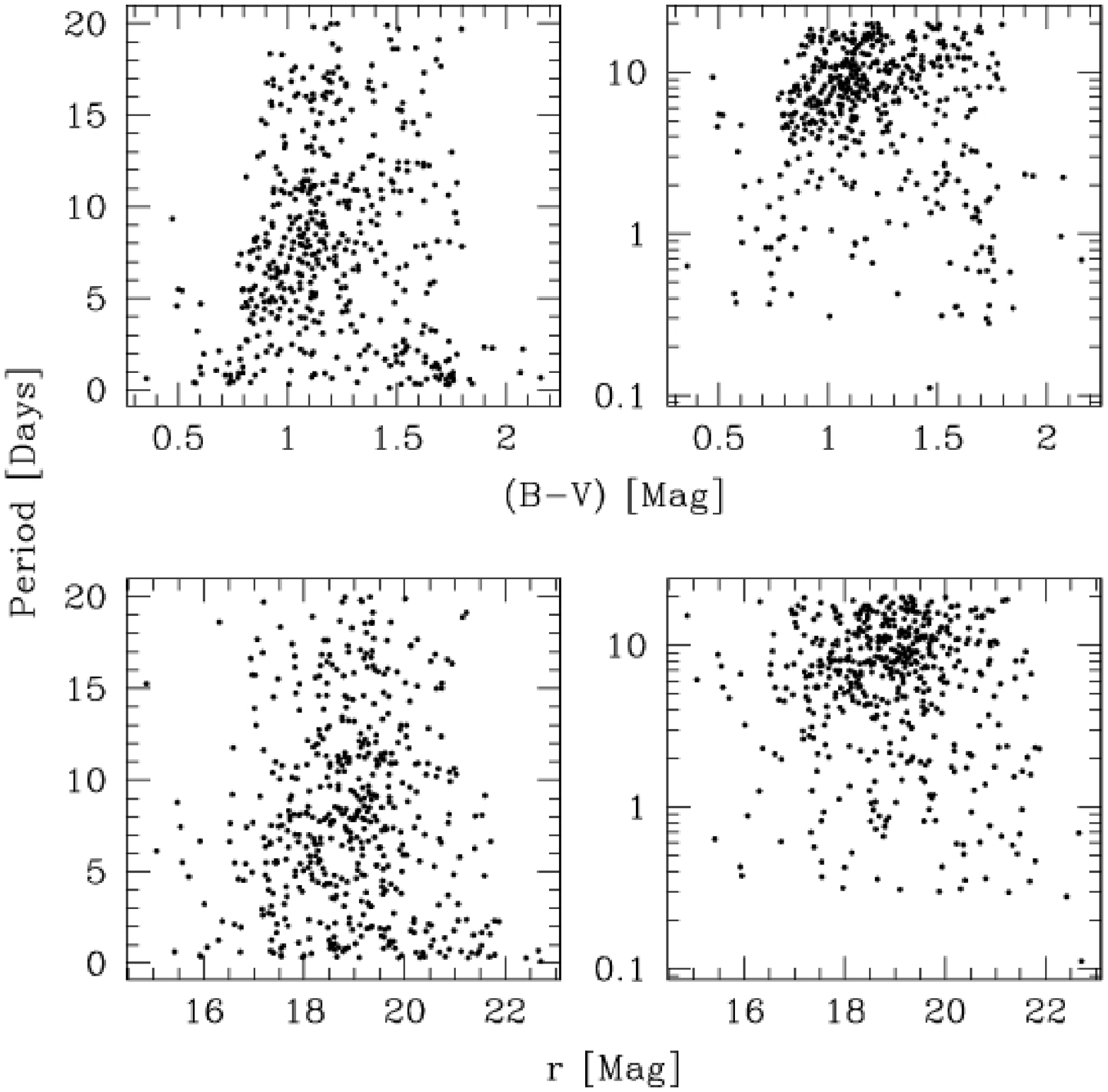}
\caption{$r$-magnitude and $(B-V)$ vs. Period for variable field stars located away from the cluster main sequence on a CMD. We exclude eclipsing binaries and pulsating variables from this plot. The period is shown on a linear scale on the left-hand-side and on a logarithmic scale on the right-hand-side. Note that no correction for redenning has been applied to the colors.}
\label{fig:fieldrots}
\end{figure}

In addition to incompleteness, the presence of field stars may also bias the observed period-color and period-amplitude distributions. Based on spectroscopic observations of several of the candidate rotational variables that we discuss in \S~\ref{sec:speccomp} we estimate that $\sim 20\%$, or $\sim 115$ of the stars in our catalog ($\sim 74$ after applying the multi-harmonic AoV selection) are field stars. While the field star variables located away from the cluster main sequence on a CMD do not show an obvious correlation between color and rotation period (figure~\ref{fig:fieldrots}), we cannot infer that the same is true for the field stars near the cluster main sequence since these stars generally have different masses and radii. To disentangle the period-color distribution of field stars from cluster members would require either a complete spectroscopic survey of all the rotational variables, or a time series study of a field off the cluster. Note that the estimated contamination fraction is small enough that the conclusions of the paper should not be strongly affected by the contamination. 

Out of a total sample of $\sim 1450$ cluster members with $14.5 < r < 22.3$ that we have observed, we estimate that $\sim 460$ are detected as variables, and $\sim 220$ have semi-amplitudes greater than $0.01~{\rm mag}$. Note that we used a stricter $\chi^{2}$-based membership selection for choosing candidate variable cluster members than what was used to estimate the total number of surveyed cluster members in Paper I. The number of cluster members that could have potentially been selected as variable cluster members is thus likely to be slightly smaller than the estimate of $\sim 1450$. Based on the completeness estimates for spot-model injected light curves without multi-harmonic AoV selection, we estimate that we have detected $\sim 70\%$ of the large-amplitude cluster member variables in this sample, so we estimate that $\sim 20\%$ of cluster members are variable with semi-amplitudes greater than $0.01~{\rm mag}$.

\subsection{Comparison with the RACE-OC Project}

\citet{Messina.08} have recently presented rotation periods for a number of stars in M37 as part of the RACE-OC project. We find that 91 of the 106 candidate cluster member stars listed in table 2 of \citet{Messina.08} that are not classified as $\delta$-Scuti, RR Lyr or eclipsing binaries match to stars in our point source catalog. To do the matching we allow for a second-order polynomial transformation from the \citet{Messina.08} coordinates to our coordinates, and match stars within a radius of $5\arcsec$. For sources that match to multiple stars in our catalog we choose the match with the smallest magnitude difference. Thirteen of the 15 unmatched sources lie on a Megacam chip gap, while stars 2835 and 3021 do not lie within $5\arcsec$ of any stars in our point source catalog. We have independently detected periodic variability for 58 of the 91 matched stars. We classify 38 of these stars as potential cluster members, 21 are not classified as potential cluster members, while 2 stars do not have $BV$ photometry from \citet{Kalirai.01} and are excluded from the catalog of rotating candidate cluster members presented in this paper. Note that the membership classification by \citet{Messina.08} is based on photometry through two filters while our classification utilizes five filters. Of the 32 stars that we do not classify as variables, 19 are saturated in more than two-thirds of our time-series observations, 11 have a light curve RMS that is higher than the median RMS as a function of magnitude, while two have an RMS that is lower than the median RMS as a function of magnitude.

\begin{figure}[p]
\plotone{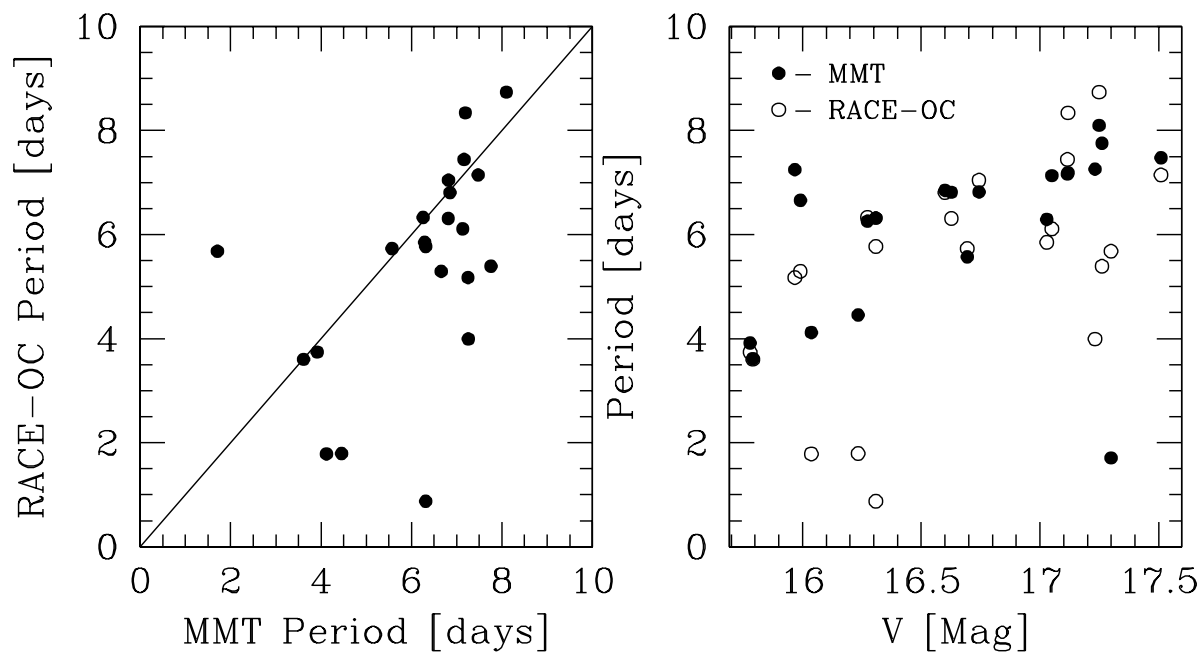}
\caption{(Left) Comparison between the periods presented in this paper (MMT Period) and the Lomb-Scargle periods measured by the RACE-OC project for the 22 matching stars included in our clean sample. Note that our periods are generally longer than the periods measured by the RACE-OC project. (Right) We show the period-magnitude relation for the matching stars using the MMT periods (filled circles) and the RACE-OC periods (open circles).}
\label{fig:MessinaPeriodComparison}
\end{figure}

In figure~\ref{fig:MessinaPeriodComparison} we compare our periods to those measured by \citet{Messina.08} for the 22 matching stars which are included in our clean sample, we also compare the resulting period-magnitude diagrams. Note that the \citet{Messina.08} periods are generally shorter than the periods that we measure. The period measurements disagree by more than 20\% for the following 8 stars: V424 (M3208), V589 (M3245), V827 (M2257), V888 (M2395), V961 (M3866), V1008 (M2549), V1025 (M2895), V1135 (M4134), where the identification listed in parentheses for each star is taken from \citet{Messina.08}. In each case we find that our light curve does not phase at the period reported by \citet{Messina.08}, while inspection of the Scargle-periodograms displayed in figures 17-26 of \citet{Messina.08} reveals additional peaks near the periods that we have measured. Note that our light curves contain an order of magnitude more observations obtained on more than twice as many nights as the light curves used by \citet{Messina.08}, as a result the periods presented here are less susceptable to aliasing.

\section{The Period-Color Sequence in M37}

The relation between period and color for the clean sample of M37 stars seen in figure~\ref{fig:BVPeriod_multiharm} is similar to that seen for stars in the Hyades \citep{Radick.87}. As we will discuss in \S 8, this sequence provides a powerful test for theories of stellar angular momentum evolution. In this section we provide an analytic fit to the period-color relation, and also evaluate the spread in rotation periods about this fit as a function of mass.

\begin{figure}[p]
\plotone{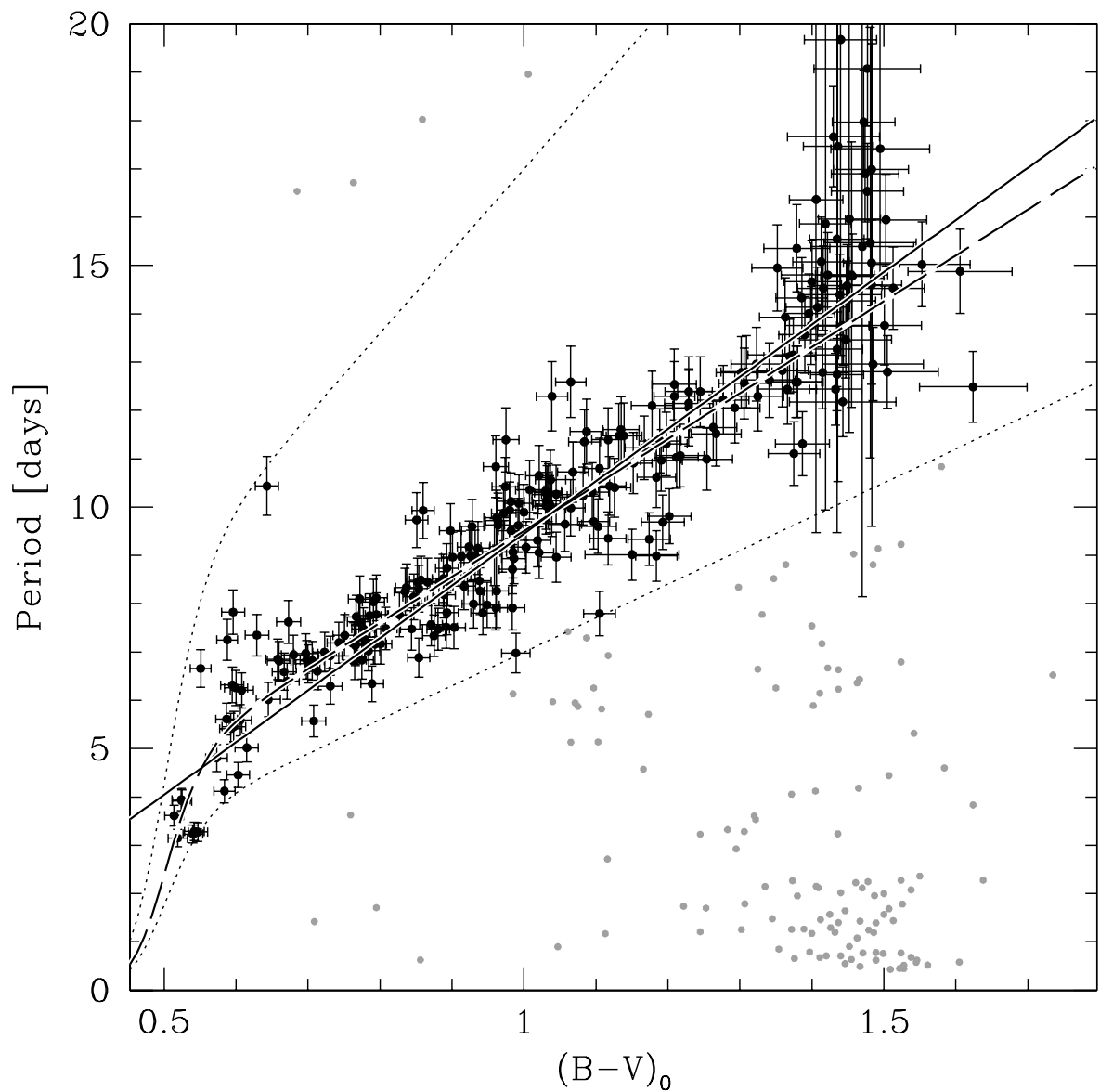}
\caption{The selection of stars on the main period-color sequence in M37. Dark points are used to plot stars in the sequence, while light points plot stars not on the sequence. The dotted lines show the boundaries for this selection. The solid and dashed lines show a linear and broken power-law fit to the sequence respectively (equations~\ref{eqn:M37sequence_linear}~and~\ref{eqn:M37sequence_bpl}).}
\label{fig:M37sequence}
\end{figure}

The selection of stars in the $(B-V)_{0}$-period sequence is shown in figure~\ref{fig:M37sequence}. We take the period uncertainty for each star to be the quadrature sum of the bootstrap error for the star, $0.05\%$ to account for errors in the model, and $0.03\%$ to account for differential rotation. The color error for each star is interpolated from table~\ref{tab:coloruncertainties}. We fit two models of the form:
\begin{equation}
P = a_{lin} (B-V)_{0} + b_{lin}
\label{eqn:M37sequence_linear}
\end{equation}
and
\begin{equation}
P = \frac{a_{bpl}}{\left( \frac{(B-V)_{0}}{0.5} \right) ^{b_{bpl}} + \left( \frac{(B-V)_{0}}{0.5} \right) ^{-1}}
\label{eqn:M37sequence_bpl}
\end{equation}
by minimizing the total $\chi^{2}_{tot}$
\begin{equation}
\chi^{2}_{tot} = \sum_{i}\left( \left(\frac{x_{i} - x_{i,0}}{\sigma_{x,i}} \right) ^{2} + \left(\frac{y(x_{i}) - y_{i,0}}{\sigma_{y,i}} \right) ^{2} \right)
\end{equation}
where $x_{i,0}$ is the observed $(B-V)_{0}$ value of star $i$, $x_{i}$ is the predicted $(B-V)_{0}$ value for star $i$ and is treated as a free parameter, $\sigma_{x,i}$ is the uncertainty in $(B-V)_{0}$ for star $i$, and the $y$ values correspond to periods with $y(x_{i})$ being given by equation~\ref{eqn:M37sequence_linear}~or~\ref{eqn:M37sequence_bpl} for the free $a$ and $b$ parameters.

We use the downhill simplex algorithm \citep{Nelder.65,Press.92} to fit each relation solving for the $a$ and $b$ parameters as well as the $x_{i}$ values for each star. Note that this is equivalent to minimizing the orthogonal $\chi^{2}$ distance between each point and the model. The resulting parameters with uncertainties are listed in table~\ref{tab:M37sequencefit}. We list the $1\sigma$ uncertainties from 1000 bootstrap simulations (i.e. simulating data sets by resampling with replacement from the original data set), and from 1000 Monte Carlo simulations (i.e. simulating data sets by adjusting each observed $x$ and $y$ value by random variables drawn from normal distributions with standard deviations $\sigma_{x}$ and $\sigma_{y}$). We find that $\chi^{2}$ per degree of freedom for the linear and broken power-law relations are given respectively by $\chi^{2}_{dof,lin} = 3.23$ and $\chi^{2}_{dof,bpl} = 2.48$, where there are 242 degrees of freedom. We note, therefore, that we do detect a significant spread in rotation period about the main period-color sequence beyond what is due to observational uncertainties in the period and color (at the $16\sigma$ level for the broken power-law). This can be seen both from the deviation of $\chi^{2}_{dof}$ from $1$ and from the fact that the parameter errors from the bootstrap simulations are consistently larger than the parameter errors from the Monte Carlo simulations. In table~\ref{tab:M37sequence} we list the residual $RMS$ and $\chi^{2}$ per degree of freedom in $(B-V)_{0}$ bins for both the linear and broken power-law relations. Note that the spread in period is significant at greater than the $3\sigma$ level for all color bins with $(B-V)_{0} < 1.5$.

\section{Comparison With Spectroscopy}\label{sec:speccomp}

Of the 127 stars for which we obtained spectra with Hectochelle, 41 match to candidate rotational variables. Of these, 33 have an average radial velocity that is within $3\sigma$ of the cluster systemic radial velocity (see Paper I).

\begin{figure}[p]
\plotone{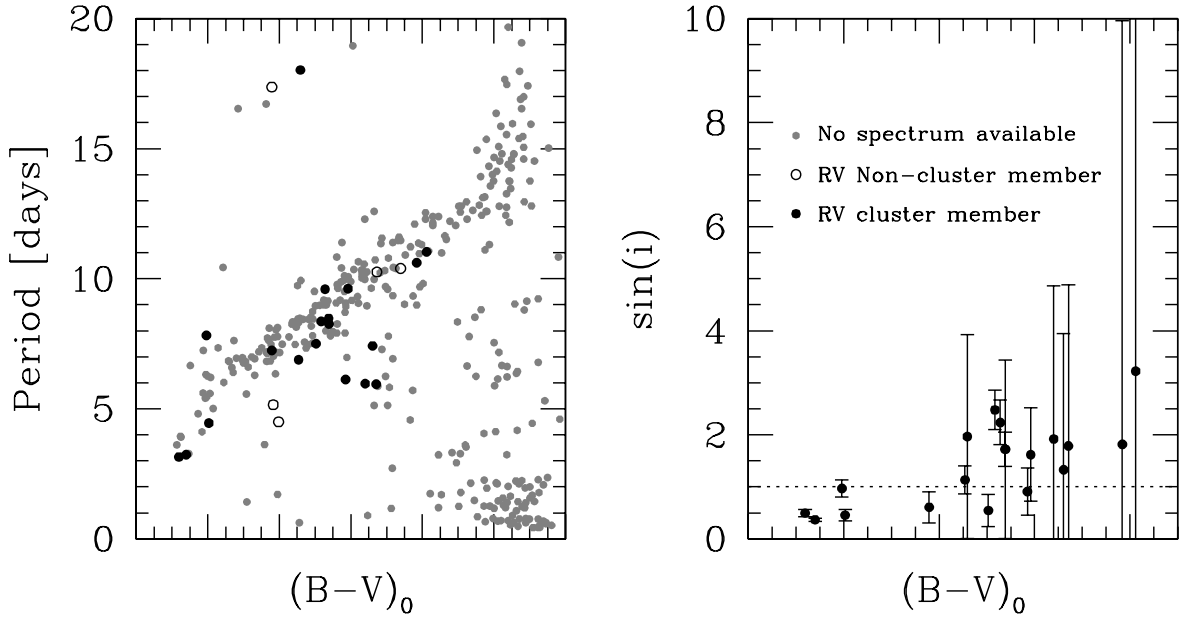}
\caption{(Left) Rotation period versus $(B-V)_{0}$ for stars in the clean sample with and without spectroscopy. (Right) $\sin i$ determined via equation~\ref{eqn:sini} versus $(B-V)_{0}$ for 19 rotational variables in the clean sample with Hectochelle spectra that have an average radial velocity consistent with being members of the cluster. The errors are dominated by uncertainties in $v \sin i$, which are estimated from the scatter in $v \sin i$ measured on four different nights. The dotted line shows $\sin i = 1$, points below this line have consistent $v\sin i$, period and radius values.}
\label{fig:BVPer_withspec}
\end{figure}

In figure~\ref{fig:BVPer_withspec} we show the stars with spectroscopy on the $(B-V)_{0}$-period relation. Using the measured $v\sin i$ and rotation periods, together with the stellar radii inferred from the best fit YREC isochrone, we can estimate the inclination angle of the rotation axis via:
\begin{equation}
\sin i = \frac{P v \sin i}{2 \pi R}.
\label{eqn:sini}
\end{equation}
In figure~\ref{fig:BVPer_withspec} we plot the sine of the angle as a function of $(B-V)_{0}$ for the 33 stars with RV consistent with being cluster members. 

For all but three stars in the clean sample the measured values of $P$ and $v \sin i$ appear to be consistent with the inferred radii since we find values that are consistent with $\sin i < 1$. The errors on this determination are dominated by the errors on $v \sin i$.

\section{Amplitude Distribution}

Because photometric observations of spotted stars are only sensitive to changes in the flux integrated over the visible hemisphere of the star, much of the information on the actual surface brightness distribution is lost. Due to well known degeneracies between the latitude, area, and temperature of a spot, it is not generally possible to obtain a unique fit to a single-filter light curve using a simple single spot model \citep[e.g.][]{Dorren.87}. When multiple spots are present their individual signals merge into a rather featureless spot-wave. Nonetheless, by studying an ensemble of stars it may be possible to gain insight on the activity-rotation relation from our data. Moreover, from an observational point of view, knowing the distribution of amplitudes is useful for planning purposes since the amplitude and period will determine if the rotation period of a star can be measured. 

We calculate the amplitudes for the clean sample of 372 stars using equation~\ref{eqn:fourier} with $N = 2$. We take the amplitude ($A_{r}$) to be the peak-to-peak amplitude of the Fourier series.

\begin{figure}
\plotone{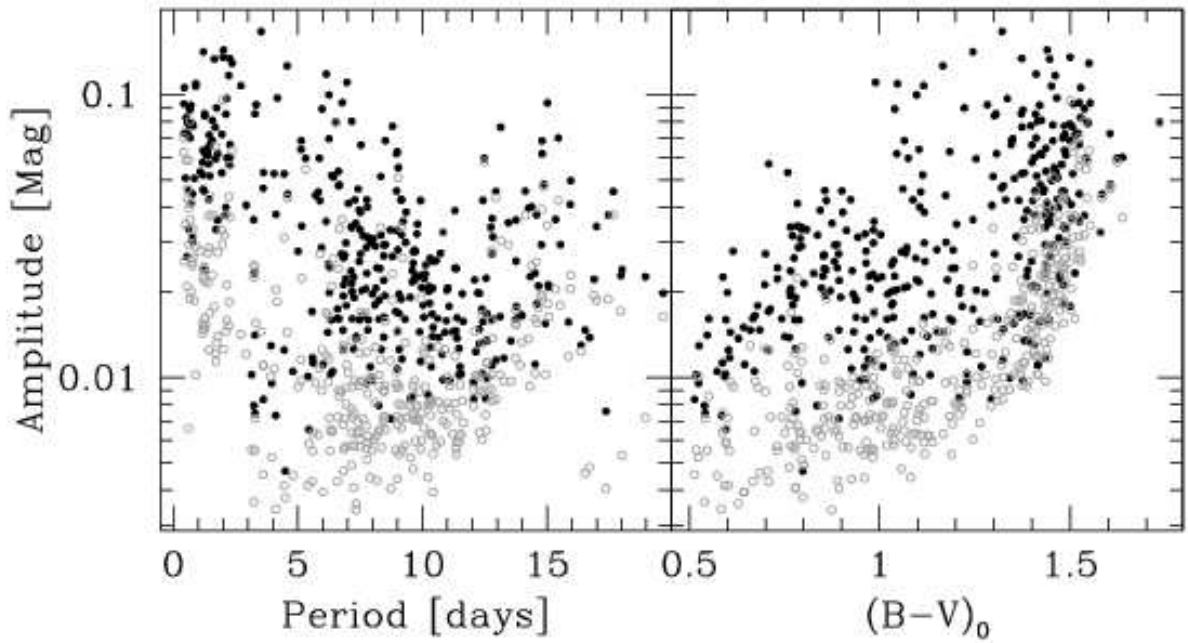}
\caption{The amplitude of the rotational variables is plotted against period and $(B-V)_{0}$. The light colored open circles show the minimum amplitude that each star could have had and still have been detected.}
\label{fig:AmpPerBV}
\end{figure}

As seen in figure~\ref{fig:AmpPerBV} the amplitude appears to be anti-correlated with the period and positively correlated with the $(B-V)_{0}$ color. There is a fairly steep selection effect, however, between the amplitude and $(B-V)_{0}$ which is caused by the drop in photometric precision for fainter stars. Due to the non-trivial relation between period and color, the selection in the period-amplitude plane is complicated. 

To evaluate whether or not the observed correlations are due to selection effects we determine for each light curve the minimum amplitude that the signal could have had and still have been detected. To do this, we find $\alpha$ such that
\begin{equation}
LS(r(t) - (1-\alpha)\tilde{r}(t),P) = -150
\label{eqn:minamp}
\end{equation}
where $LS(x,P)$ is the logarithm of the Lomb-Scargle (L-S) formal false alarm probability for light curve $x$ with period $P$ \citep[see][]{Press.92}, $\tilde{r}(t)$ is the model signal in eq.~\ref{eqn:fourier}, $r(t)$ is the observed light curve, and $-150$ is the selection threshold used to select variables in Paper II. For the purposes of this investigation we do not include stars that were not selected with L-S; we also reject stars for which no positive $\alpha$ can be found that satisfies eq.~\ref{eqn:minamp} as these are light curves for which the simple model in eq.~\ref{eqn:fourier} does not adequately describe the periodic signal. We then take the minimum observable amplitude to be:
\begin{equation}
A_{min} = \alpha A_{r}
\end{equation}
Note that we have ignored the by-eye selection that light curves are passed through following the selection on L-S. We also caution that stars with light curves that are not well modeled by the simple Fourier series will have minimum observable amplitudes that are underestimated. The minimum observable amplitudes for each point are also shown in figure~\ref{fig:AmpPerBV}.

To evaluate the significance of the apparent correlation in the presence of the selection we use Kendell's $\tau$ (non-parametric correlation statistic) modified for the case of data suffering a one-sided truncation \citep{Tsai.90,Efron.92,Efron.99}. Letting the data set be represented by the set of points $\{(x_{i},y_{i},y_{min,i})\}$, where $y_{min,i}$ is the minimum value of $y_{i}$ that could have been measured for observation $i$, define the set of comparable pairs
\begin{equation}
\mathcal{C} = \{(i,j)\,|\,y_{i} \geq y_{min,j}\ {\rm and}\ y_{j} \geq y_{min,i}\}.
\label{eq:defC}
\end{equation}
Define the risk-set numbers by
\begin{equation}
N_j = \#\{i\,|\,y_{min,i} \leq y_{j}\ {\rm and}\ y_{i} \geq y_{j}\}.
\label{eqn:risksetnumber}
\end{equation} 
The normalized correlation statistic is then
\begin{equation}
\hat{T} = \left( \sum_{(i,j) \in \mathcal{C}} {\rm sign}((x_{i} - x_{j})(y_{i} - y_{j})) \right)/\sigma
\label{eqn:defT}
\end{equation}
with the variance of the numerator given by
\begin{equation}
\sigma^{2} \approx 4\sum_{i}\frac{N_{i}^{2} - 1}{12}.
\end{equation}
The statistic is normalized such that a value of $\hat{T} = 1$ corresponds to a $1\sigma$ rejection of the null hypothesis of non-correlation.

Applying eq~\ref{eqn:defT} to the $(P,A_{r})$ and $((B-V)_{0},A_{r})$ data we find $\hat{T} = -11091$ and $\hat{T} = 3900$ respectively. We conclude, therefore, that there is a significant anti-correlation between rotation period and amplitude and a positive correlation between $(B-V)_{0}$ and amplitude for the lower main sequence stars in this cluster.

As shown by \citet{Noyes.84}, there is a tight correlation between stellar activity, measured via the ratio of emission in the cores of the Ca II H and K lines to the total luminosity of the star, and the Rossby number ($R_{O}$, the ratio of the rotation period to the characteristic time scale of convection). \citet{Messina.01a} find that the light curve amplitude also appears to be anti-correlated with $R_{O}$. 

\begin{figure}
\plotone{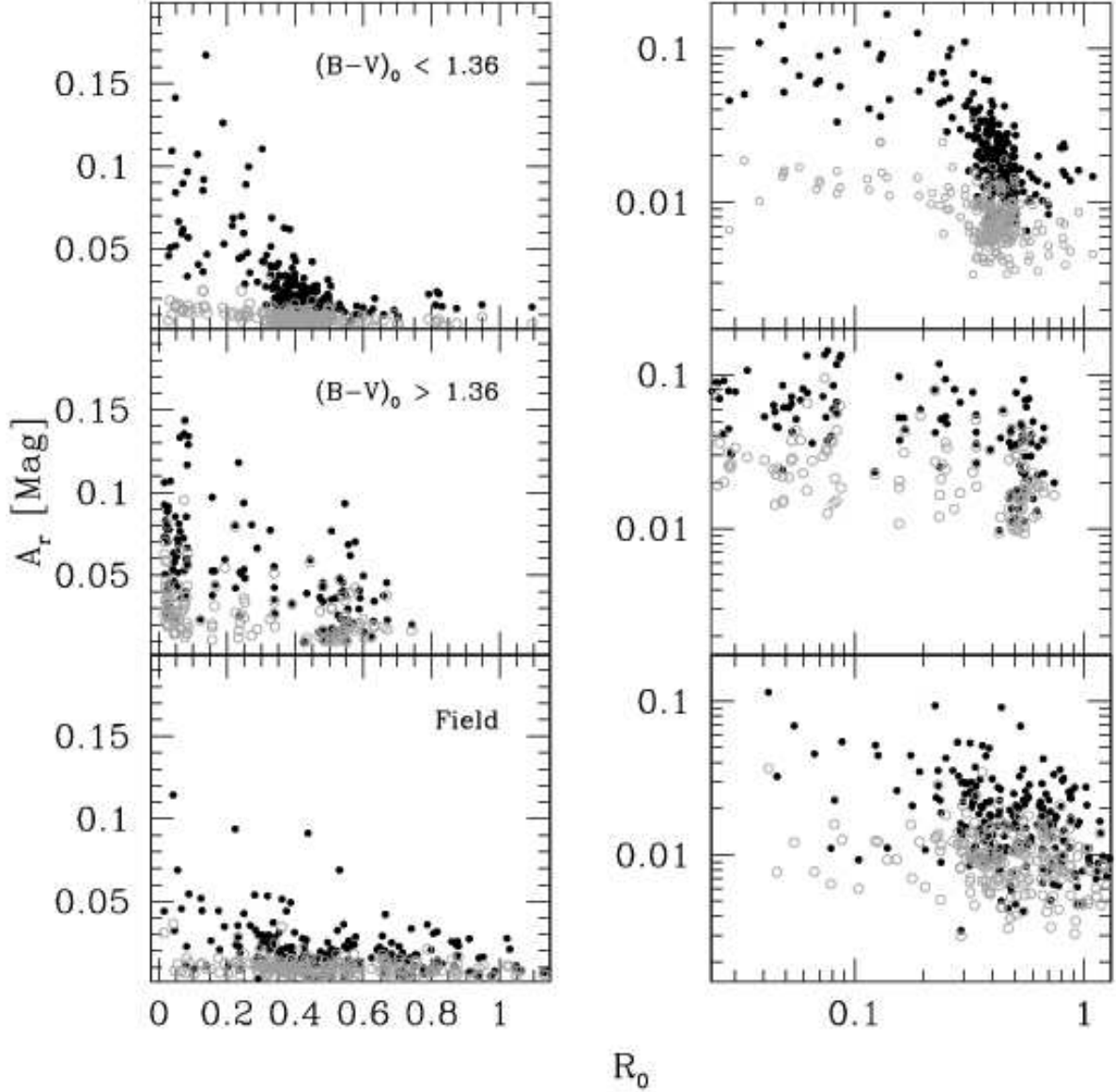}
\caption{The $r$-band peak-to-peak amplitude $A_{r}$ is plotted against the Rossby number $R_{O}$. The top two rows show the sample of clean cluster members that have $(B-V)_{0} < 1.36$ and $(B-V)_{0} > 1.36$. The bottom row shows the relation for field variables presented in Paper II, note that these stars all have $(B-V)_{0} < 1.36$. The left column is on a linear scale, the right column is on a logarithmic scale. The solid points show the measured amplitudes, the open circles show the minimum detectable amplitude for each star. Notice that for cluster members with $(B-V)_{0} < 1.36$, $R_{O}$ and $A_{r}$ are strongly anti-correlated above $R_{O} > 0.3$ while the relation flattens below $R_{O} < 0.3$. Stars with $R_{O} < 0.3$ appear to have a minimum amplitude of $A_{r} \sim 0.03~{\rm mag}$. Cluster members with $(B-V)_{0} > 1.36$ show a relatively flat relation even above $R_{O} > 0.3$. The field stars also show a relatively flat relation.}
\label{fig:AmplitudeRossby}
\end{figure}

To compute $R_{O}$ for each star we use the empirical expression for the convective time-scale from \citet{Noyes.84}
\begin{equation}
\log \tau_{c} = \left\{ \begin{array}{ll}
1.362 - 0.166x + 0.025x^{2} - 5.323x^{3}, & x > 0 \\
1.362 - 0.14x, & x < 0
\end{array}
\right.
\end{equation}
where $\tau_{c}$ is the convective time-scale in days and $x = 1.0 - (B-V)_{0}$. We plot the amplitude of the variables against $R_{O}$ in figure~\ref{fig:AmplitudeRossby}, showing stars with $(B-V)_{0} < 1.36$ and $(B-V)_{0} > 1.36$ separately. For comparison we also show the candidate field rotators from Paper II. 

\begin{figure}
\plotone{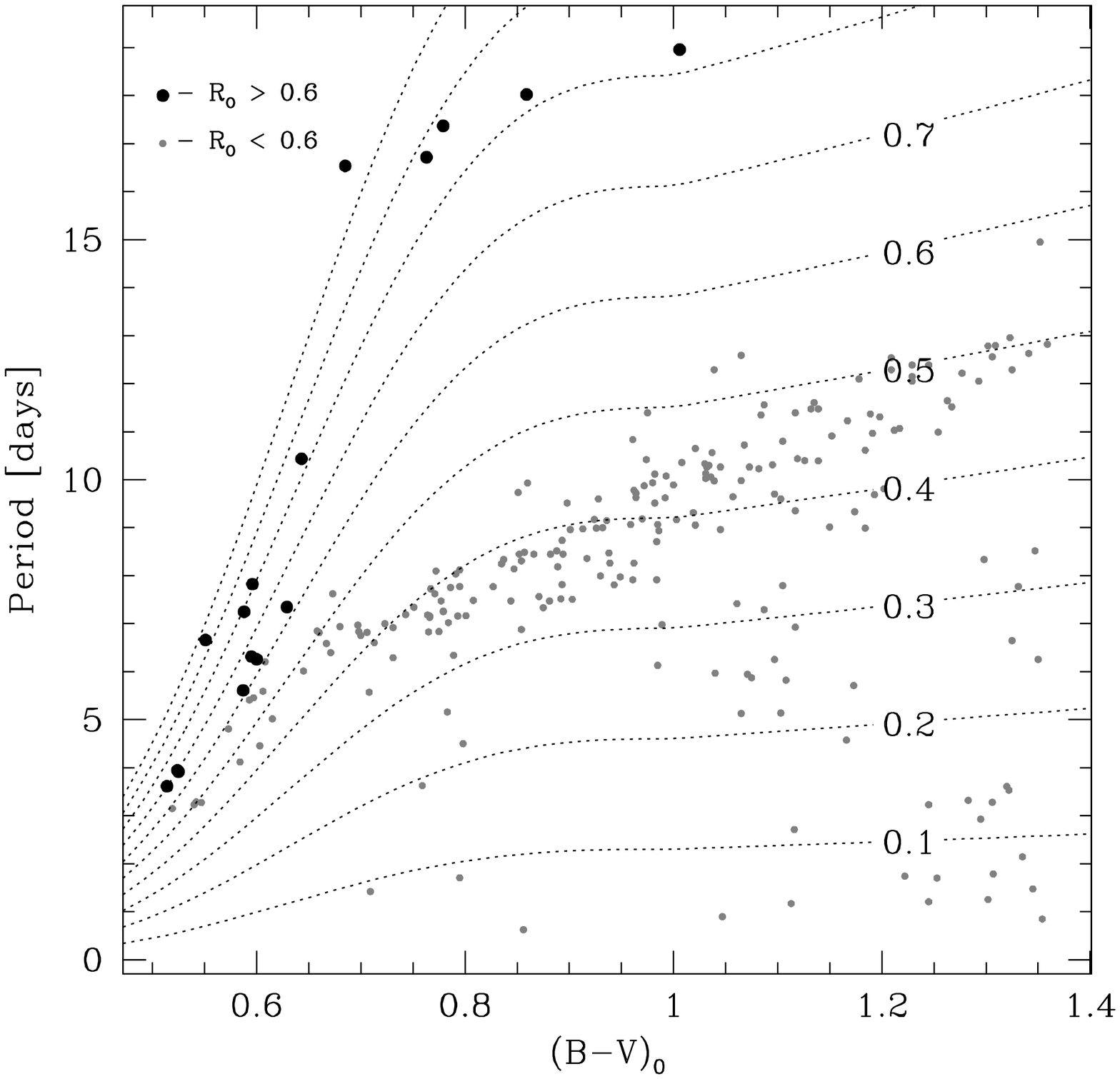}
\caption{Comparison of stars with $R_{O} < 0.6$ and $R_{O} > 0.6$ on the $(B-V)_{0}$-$P$ plot. Only stars in the clean sample with $(B-V)_{0} < 1.36$ are shown. The dotted lines are lines of constant $R_{O}$.}
\label{fig:BVPer_highRossby}
\end{figure}

The anti-correlation between $R_{O}$ and $A_{r}$ appears to be steeper for stars with $(B-V)_{0} < 1.36$ than for stars with $(B-V)_{0} > 1.36$. We note that the empirical constraints on the convective time-scale are less stringent for redder stars, so the values of $R_{O}$ are more susceptible to systematic errors for $(B-V)_{0} > 1.36$. Focusing on cluster members with $(B-V)_{0} < 1.36$, the relation between $R_{O}$ and $A_{r}$ appears to flatten out, or saturate, for stars with $R_{O} < 0.3$. There appears to be a dearth of low-amplitude stars with $R_{O} < 0.2$, note that stars with peak-to-peak amplitudes $A_{r} > 0.02~{\rm mag}$ should have been detectable. There is also a hint that the relation between $R_{O}$ and $A_{r}$ is flat for $R_{O} > 0.6$. Note that as seen in figure~\ref{fig:BVPer_highRossby} many of the stars with $R_{O} > 0.6$ fall above the main period-color sequence. As we discuss below, it is possible that the periods measured for these stars are not the rotation periods but rather the spot-evolution timescales.

Applying eq~\ref{eqn:defT} to the $(R_{O},A_{r})$ data we find $\hat{T} = -16462$ for all stars in the clean data set, $\hat{T} = -10520$ for 233 stars with $(B-V)_{0} < 1.36$ and $\hat{T} = -1150$ for 129 stars with $(B-V)_{0} > 1.36$. For comparison, the $((B-V)_{0},A_{r})$ data has $\hat{T} = 3846$ for $(B-V)_{0} < 1.36$ and $\hat{T} = -540$ for $(B-V)_{0} > 1.36$, while the $(P,A_{r})$ data has $\hat{T} = -5430$ for $(B-V)_{0} < 1.36$ and $\hat{T} = -1145$ for $(B-V)_{0} > 1.36$. For $(B-V)_{0} < 1.36$ the correlation for $(R_{O},A_{r})$ is more significant than for $((B-V)_{0},A_{r})$ or $(P,A_{r})$. 

\begin{figure}
\plotone{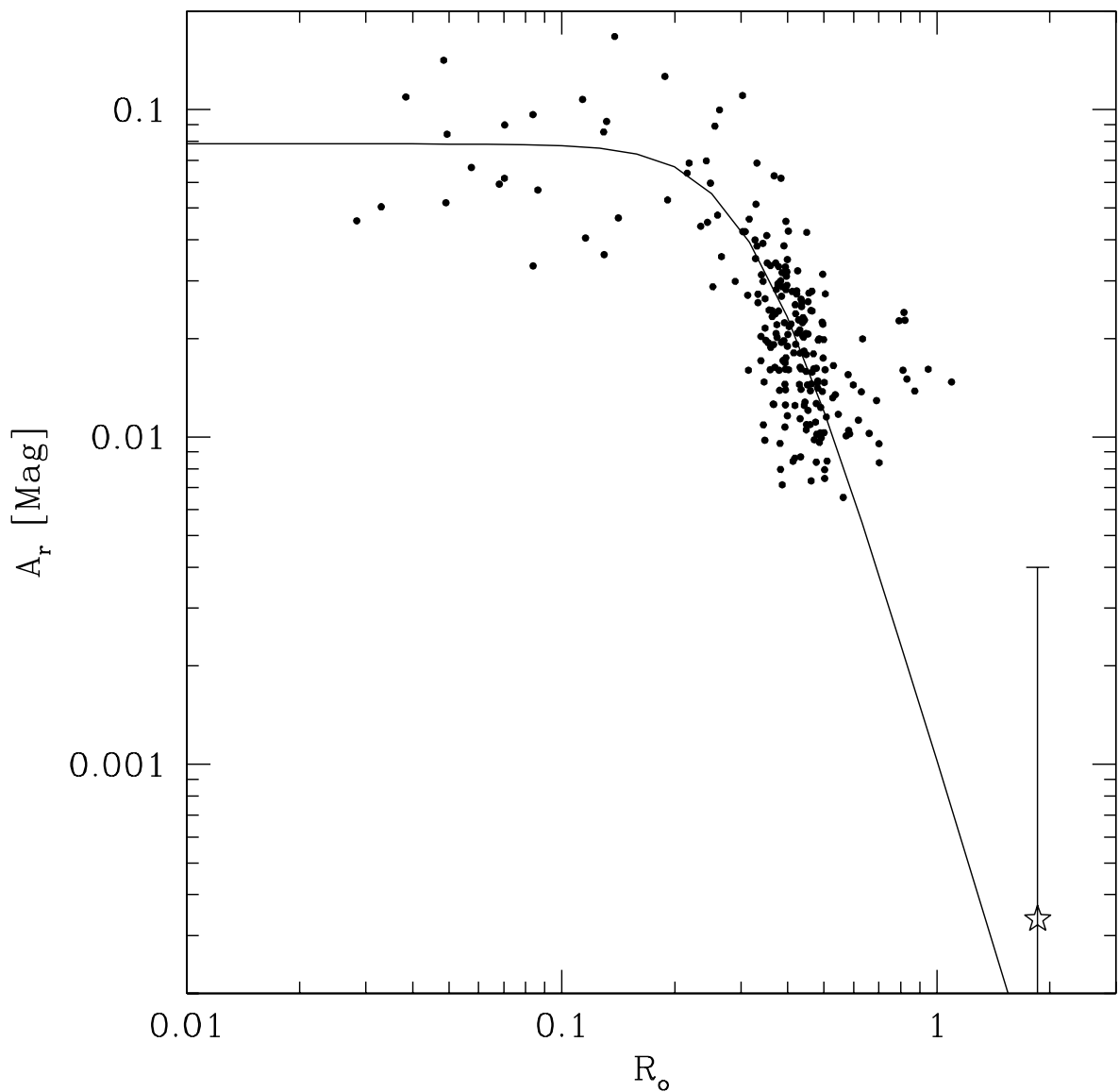}
\caption{The $r$-band peak-to-peak amplitude $A_{r}$ is plotted against the Rossby number $R_{O}$ for the sample of clean stars that were selected by L-S and have $(B-V)_{0} < 1.36$. The solid line shows equation~\ref{eqn:arrofit} which is a fit to the data with $R_{O} < 0.6$. The star shows the approximate location of the Sun for a typical large sunspot, while the errorbar indicates the approximate upper limit on the amplitude.}
\label{fig:AmplitudeRossbyFit}
\end{figure}

For $(B-V)_{0}$, $R_{O} < 0.6$ the selection on $A_{r}$ does not appear to bias the $(R_{O},A_{r})$ distribution. We find that the $(R_{O},A_{r})$ data for cluster members with $(B-V)_{0} < 1.36$ and $R_{0} < 0.6$ can be fit with the function:
\begin{equation}
A_{r} = \frac{0.078 \pm 0.008}{1 + \left( \frac{R_{O}}{0.31 \pm 0.02}\right) ^{3.5 \pm 0.5}}
\label{eqn:arrofit}
\end{equation}
where the errors listed are the $1\sigma$ errors from 1000 bootstrap simulations. Figure~\ref{fig:AmplitudeRossbyFit} shows this relation. We also show the approximate location of the Sun on this diagram, assuming a typical large sunspot with an area that is $500$ millionths that of the solar disk, a temperature ratio to the photosphere of $0.7$, the bolometric corrections for the $r$-band from \citet{Girardi.04} and $(B-V)_{\odot} = 0.656$ \citep{Gray.92} to convert the bolometric amplitude to an $r$-band amplitude, and the solar equatorial rotation period of 24.79 days \citep{Howard.84}. The error bar shows the approximate range of values for the Sun, where the upper limit is for the largest sunspot group observed to date \citep[the giant sunspot group of April, 1947 had an area of $\sim 6000$ millionths that of the solar disk and was large enough to be seen without optical aid, see][]{Taylor.89}.

\begin{figure}
\epsscale{0.9}\plotone{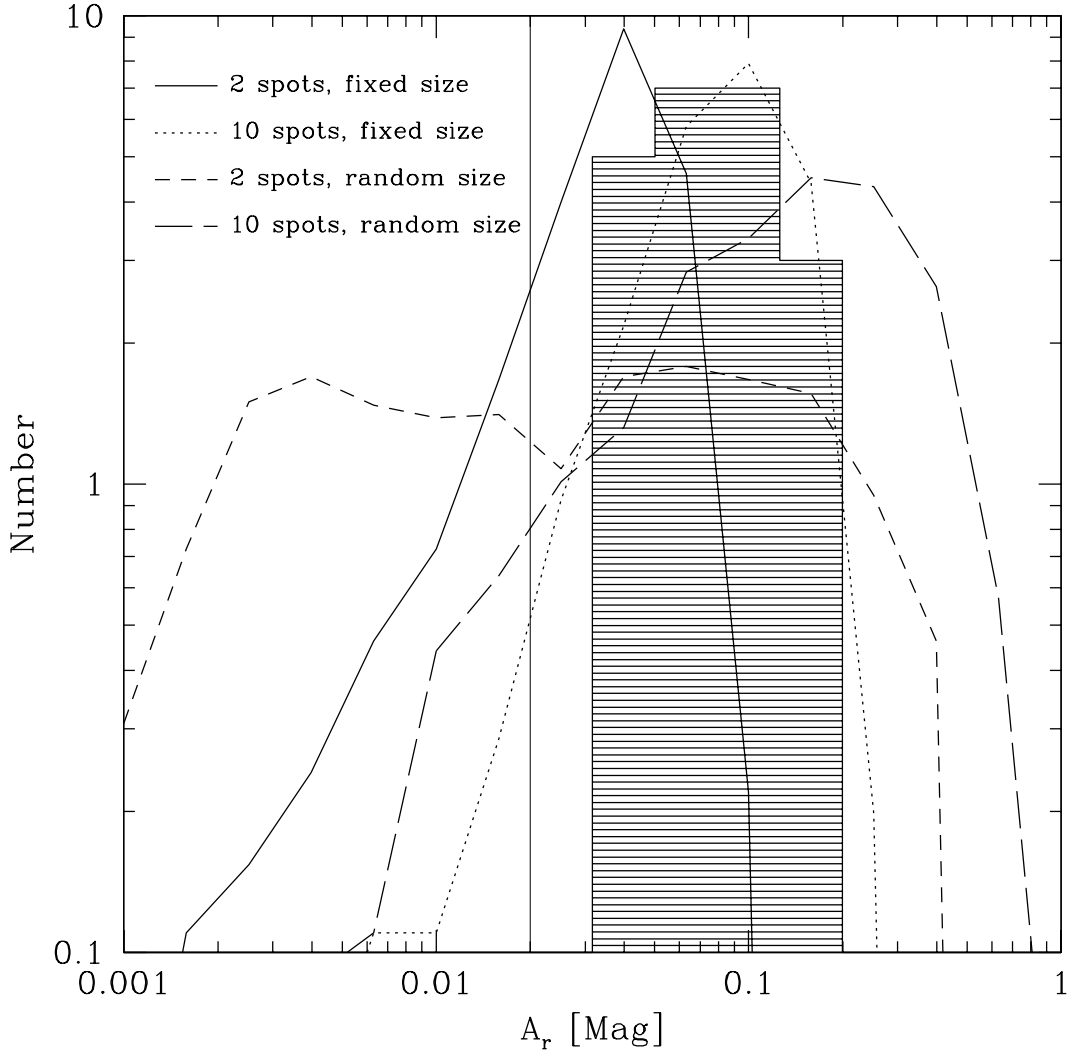}
\caption{The distribution of peak-to-peak $r$-band amplitude measurements ($A_{r}$) for stars with $(B-V)_{0} < 1.36$, $R_{O} < 0.2$ (shaded histogram) compared with model simulations. The models have been normalized to have the same total number of stars as the observed distribution. The simulations assume either a fixed spot size that is $1\%$ of the stellar surface area or allow the spot-sizes to be uniformly distributed in logarithm between $10^{-5}$ times the stellar surface area and $10\%$ of the stellar surface area. The simulations also assume either two spots per star or ten spots per star. The vertical line indicates the minimum detectable amplitude for stars in the cluster. Note that the lack of observed stars with $0.02 < A_{r} < 0.03$ indicates that the observed distribution is peaked at $A_{r} > 0.03~{\rm mag}$. Models with a fixed spot-size result in a peaked $A_{r}$ distribution whereas models that allow a broad spot-size distribution result in a broad $A_{r}$ distribution that is inconsistent with the observations. Increasing the number of spots per star can tighten up the $A_{r}$ distribution when the spot-size distribution is broad.}
\label{fig:spotamplitudedistribution}
\end{figure}

As noted above, there appears to be a minimum amplitude of $\sim 0.03~{\rm mag}$ for stars with $(B-V)_{0} < 1.36$, $R_{O} < 0.2$. The fact that there are stars with $R_{O} > 0.3$ that have amplitudes down to the minimum detectable limits below $0.01~{\rm mag}$ shows that this is not likely a result of underestimating the minimum detectable amplitude and is likely to be a real effect. This indicates that these rapidly rotating stars may have a distribution of spot-sizes that is peaked, or that they have several spots. To illustrate this we have conducted Monte Carlo simulations of spotted star light curves for four different models of the spot-size and spot-number distributions. For the spot-size distribution we either fixed the spot area to $1\%$ that of the stellar surface area or we allowed the spot-sizes to be uniformly distributed in logarithm between $10^{-5}$ times the stellar surface area and $10\%$ of the stellar surface area. We calculated models with two spots per star or with ten spots per star. For each model we simulate 1000 light curves using the latest version of the Wilson-Devinney program \citep{Wilson.71, vanHamme.03}. The simulations are conducted assuming a spot to photosphere temperature ratio of 0.7, random orientations for the rotation axis, and that the spots are distributed randomly over the surface of the star. Figure~\ref{fig:spotamplitudedistribution} shows the distribution of $A_{r}$ for the model simulations together with the observed $A_{r}$ distribution for stars with $(B-V)_{0} < 1.36$, $R_{O} < 0.3$. While fitting a model to the observed distribution is beyond the scope of this paper, we note that the models with a fixed spot-size or ten spots per star have peaked $A_{r}$ distributions whereas the model with a broad spot-size distribution and only two spots per star results in a broad $A_{r}$ distribution that is inconsistent with our observations. This suggests that the logarithmic distribution of spot filling-factors (i.e. the fraction of the stellar surface that is covered by spots) is peaked.

\section{The Blue K Dwarf Phenomenon in M37}

It is well known that the K dwarfs in the Pleiades fall blueward of a main sequence isochrone when plotted on a $B-V$ CMD \citep[see][hereafter S03]{Stauffer.03}. S03 have argued that this is due to differences in the spectral energy distribution of the Pleiades stars and the field dwarfs which are used to define the main sequence. They argue that this difference is due to more significant cool spots and plage areas on the photospheres of the young, rapidly rotating, heavily spotted Pleiades stars than are present on the older, slowly rotating, less heavily spotted field dwarfs. The plage areas result in excess flux in the $B$ and $V$ bands, while spots cause excess flux in the near infrared. As evidence for this explanation they show that the discrepancy at fixed color between the magnitude of the Pleiades stars and the main sequence isochrone correlates with $v \sin i$. \citet[][hereafter A07]{An.07} have also shown that the K dwarfs in the young cluster NGC 2516 are too blue in $B-V$ and that the discrepancy correlates with $v \sin i$.

\begin{figure}
\plotone{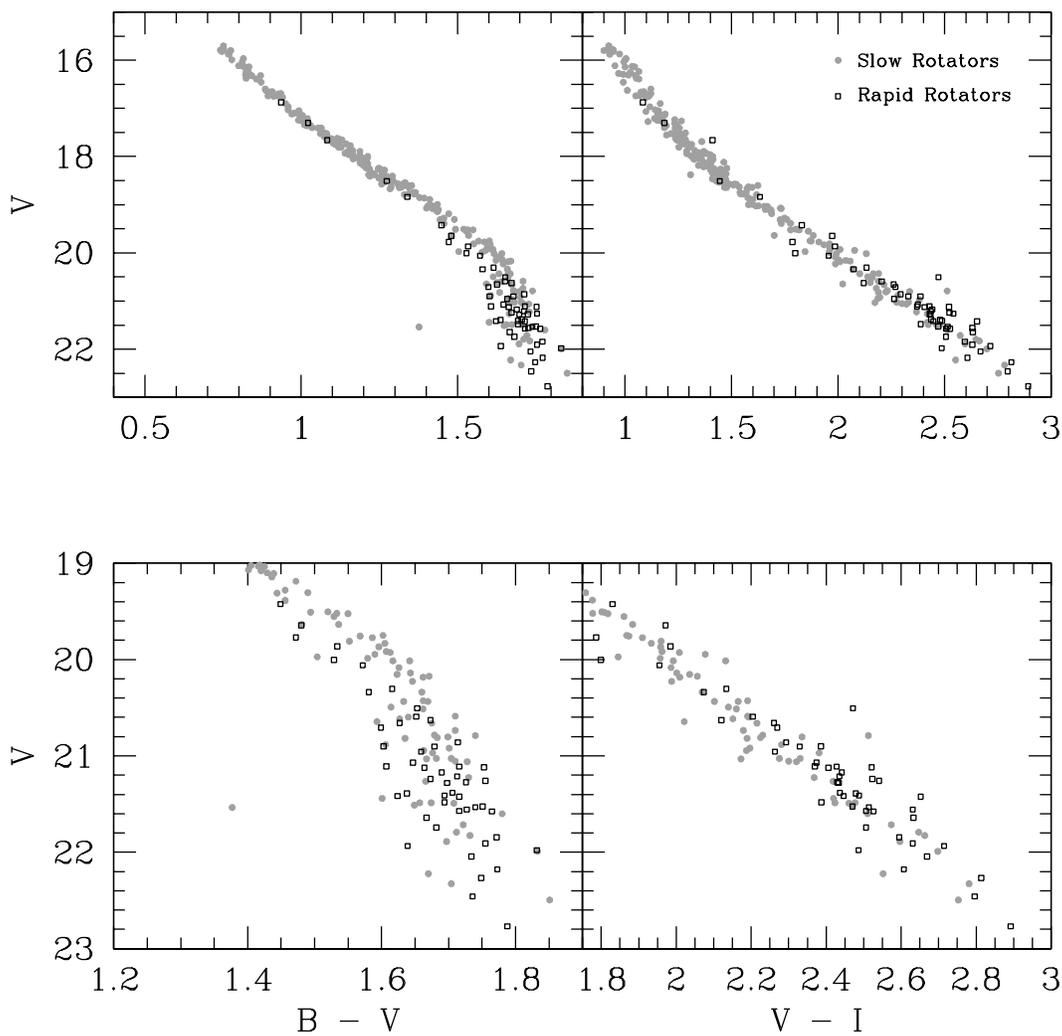}
\caption{Stars located on the main color-rotation period band (filled, light points) and rapidly rotating stars with $P < 2$ days are plotted on $B-V$ and $V-I_{C}$ CMDs. The bottom panels show a close up on the lower main sequence. Along the lower main sequence the rapidly rotating stars appear to be slightly bluer in $(B-V)$ at given $V$ and slightly redder in $(V-I_{C})$ at given $V$ than the slower rotators.}
\label{fig:BVrapidvsslow}
\end{figure}

\begin{figure}
\plotone{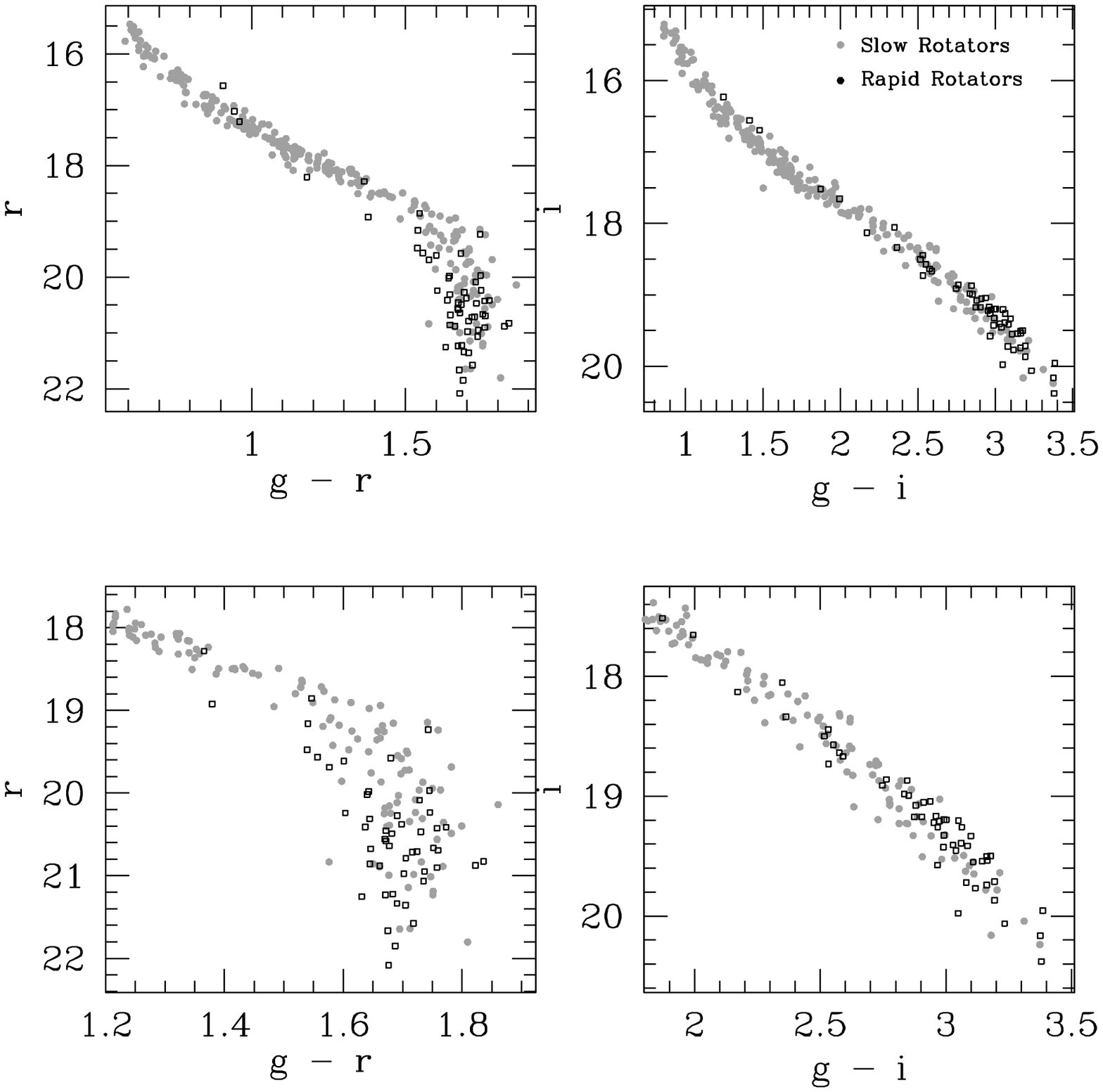}
\caption{Same as figure~\ref{fig:BVrapidvsslow} here we show $g-r$ and $g-i$ CMDs.}
\label{fig:grirapidvsslow}
\end{figure}

Using our rotation periods we can look for this phenomenon in M37. We first define two groups of stars: those which lie on the main color-rotation period band (with the selection performed as in \S 4), and rapid rotators with $P < 2$ days. In figure~\ref{fig:BVrapidvsslow} we show the locations of these two groups of stars on $B-V$ and $V-I_{C}$ CMDs and in figure~\ref{fig:grirapidvsslow} we show them on $g-r$ and $g-i$ CMDs. The results appear to be similar to that found by S03 and A07. In the $B-V$ and $g-r$ CMDs the rapid rotators appear to be bluer than the slower rotators at a fixed magnitude along the lower main sequence. While in the $V-I_{C}$ and $g-i$ CMDs the rapid rotators are slightly redder at fixed magnitude than the slower rotators.

\begin{figure}
\plotone{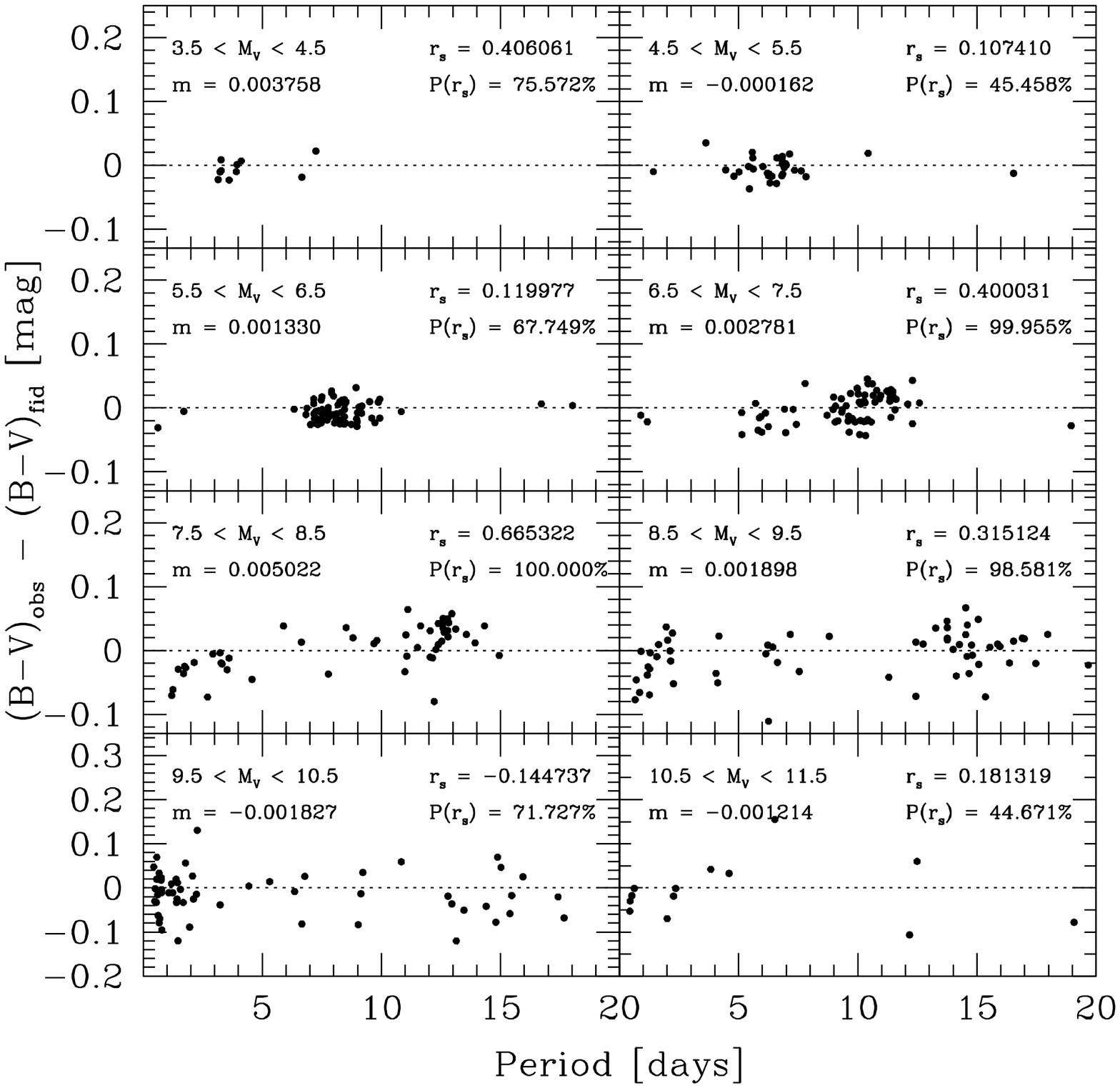}
\caption{Period versus $(B-V)_{obs} - (B-V)_{fid}$, where $(B-V)_{fid}$ is the color on the fiducial main sequence at the $V$ magnitude of each star. This is plotted for stars in different $M_{V}$ bins (assuming a distance modulus of $(m-M)_{V} = 11.572$ from Paper I). In each bin we calculate the Spearman rank-order correlation coefficient ($r_{s}$) as well as the two-sided significance level of its deviation from zero ($P(r_{s})$); we also fit a linear relation of the form $(B-V)_{obs} - (B-V)_{fid} = mP + b$ listing the slope in each bin. Stars in the magnitude bins ($6.5 < M_{V} < 9.5$) show a significant correlation between $P$ and $(B-V)_{obs} - (B-V)_{fid}$.}
\label{fig:BVcolorDiff_vsPeriod}
\end{figure}

\begin{figure}
\plotone{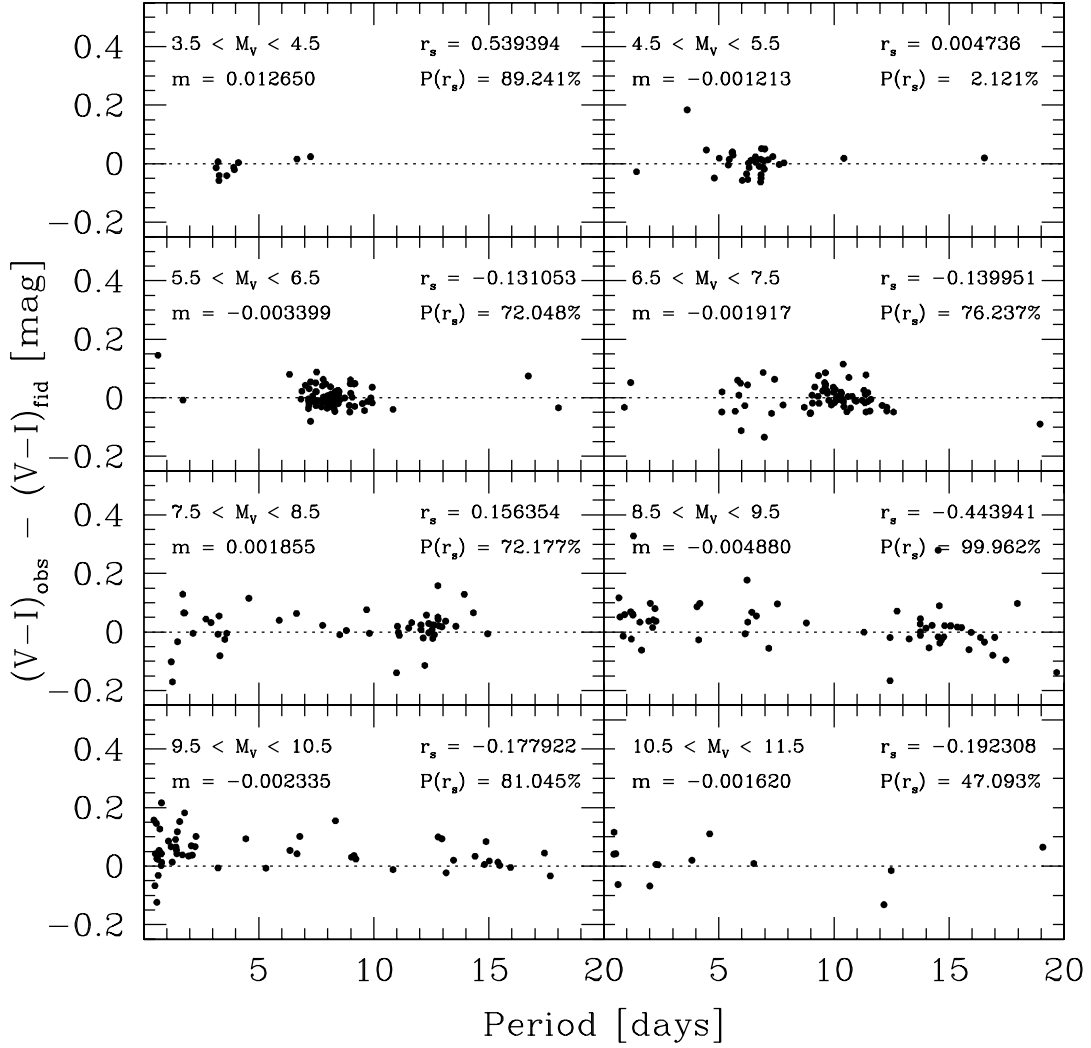}
\caption{Same as figure~\ref{fig:BVcolorDiff_vsPeriod}, here $(V-I_{C})_{fid}$ is calculated at the $V$ magnitude of the stars. In this case stars with $8.5 < M_{V} < 10.5$ show an anti-correlation between $P$ and $(V-I_{C})_{obs} - (V-I_{C})_{fid}$.}
\label{fig:VIcolorDiff_vsPeriod}
\end{figure}

\begin{figure}
\plotone{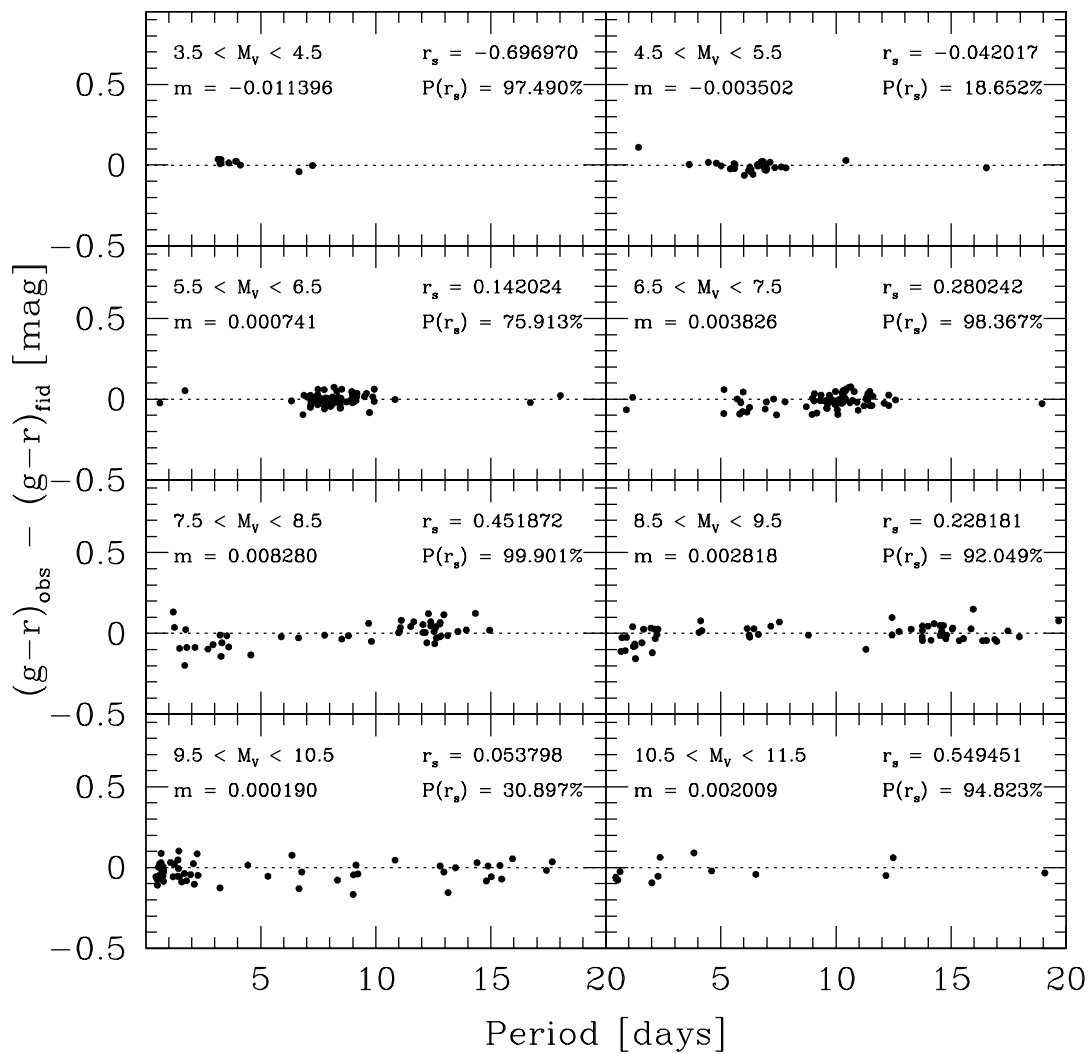}
\caption{Same as figure~\ref{fig:BVcolorDiff_vsPeriod}, this time using $g-r$. The stars with $6.5 < M_{V} < 9.5$ or $10.5 < M_{V} < 11.5$ show a significant correlation between $P$ and $(g - r)_{obs} - (g-r)_{fid}$.}
\label{fig:grcolorDiff_vsPeriod}
\end{figure}

\begin{figure}
\plotone{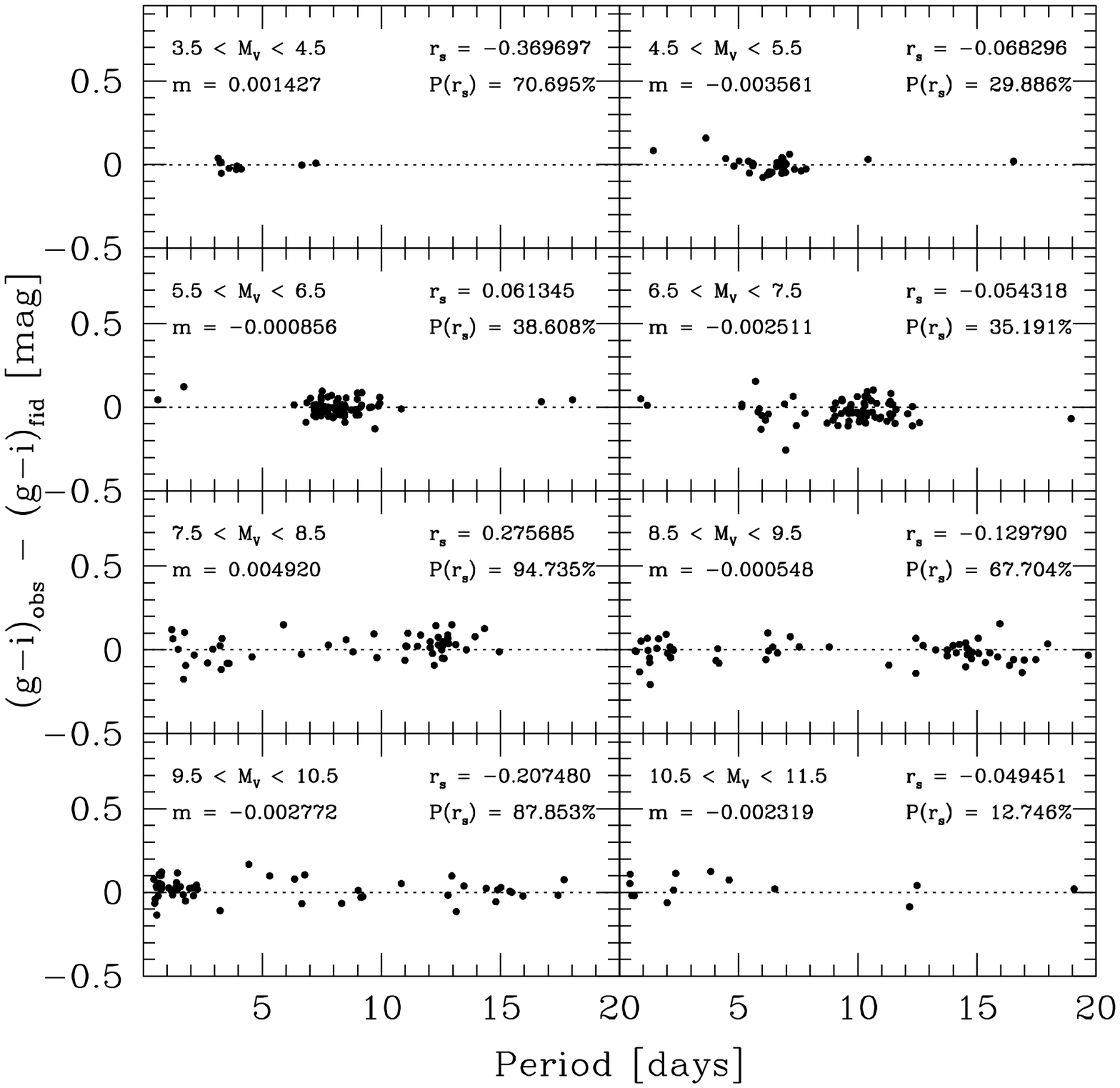}
\caption{Same as figure~\ref{fig:BVcolorDiff_vsPeriod}, this time using $g-i$. Stars with $7.5 < M_{V} < 8.5$ show a positive correlation between $P$ and $(g-i)_{obs} - (g-i)_{fid}$ while stars with $9.5 < M_{V} < 10.5$ show an anti-correlation.}
\label{fig:gicolorDiff_vsPeriod}
\end{figure}

To illustrate these effects quantitatively we bin the data by magnitude and plot in figures~\ref{fig:BVcolorDiff_vsPeriod}-\ref{fig:gicolorDiff_vsPeriod} the rotation period of the variables against the color difference $c_{obs} - c_{fid}$ where $c_{fid}$ is the color interpolated within a fiducial main sequence, at the $V$ magnitude of the star; the fiducial sequence is drawn by eye. We use a fiducial sequence rather than a theoretical sequence because the discrepancy between the observed and theoretical sequences varies with magnitude, and correlations between period and magnitude could lead to spurious correlations between period and color difference. For each bin we calculate the Spearman rank-order correlation coefficient ($r_{s}$) as well as the two-sided significance level of its deviation from zero, we ignore uncertainties in the photometry in calculating this statistic. 

As seen in figure~\ref{fig:BVcolorDiff_vsPeriod}, there is a significant correlation (greater than $98\%$ significance) between period and $(B-V)_{obs} - (B-V)_{fid}$ at fixed $V$ for stars with $6.5 < M_{V} < 9.5$. Fitting a linear relation of the form $(B-V)_{obs} - (B-V)_{fid} = mP + b$ gives a slope of $m \sim 0.005$ for the bin with the most significant correlation detection. When using $(V-I_{C})_{0}$ we find anti-correlations with greater than $80\%$ significance between period and $(V-I_{C})_{obs} - (V-I_{C})_{fid}$ for stars with $8.5 < M_{V} < 10.5$ (figure~\ref{fig:VIcolorDiff_vsPeriod}). In $g-r$, the stars with $6.5 < M_{V} < 9.5$ or $10.5 < M_{V} < 11.5$ show a positive correlation between period and $(g - r)_{obs} - (g - r)_{fid}$ (figure~\ref{fig:grcolorDiff_vsPeriod}). Finally in $g-i$, stars with $7.5 < M_{V} < 8.5$ have a positive correlation between period and $(g-i)_{obs} - (g-i)_{fid}$ (figure~\ref{fig:gicolorDiff_vsPeriod}) while stars with $9.5 < M_{V} < 10.5$ show a negative correlation.

The positive correlations seen in $B-V$ and $g-r$, together with the anti-correlations in $V-I_{C}$ and $g-i$ for the red stars is consistent with the blue K dwarf phenomenon in the Pleiades described by S03. This effect has been seen in the Pleiades and in NGC 2516 amongst stars with $0.8 < (B-V)_{0} < 1.2$ ($6 \la M_{V} \la 7.5$; see A07). We find that this phenomenon extends into the M dwarf regime where $(B-V)_{0}$ saturates at $\sim 1.5$ mag ($M_{V} \ga 9.5$).

These observations bear on a well-known problem in the theory of low-mass stars. In recent years studies of M-dwarf eclipsing binaries have revealed that these stars have radii that are $\sim 10\%$ larger than predicted by theory \citep{Torres.02, Ribas.03, Lopez-Morales.05, Ribas.06, Torres.07}. Note that the \citet{Torres.07} finding is for an M-dwarf orbited by a short-period transiting Neptune-mass planet. The observed luminosities of these stars are in good agreement with theoretical predictions, but their effective temperatures are lower than predicted. The observed low-mass eclipsing binary systems generally have short periods, and it has been suggested that the discrepancy between theory and observations may be due to enhanced magnetic activity on these rapidly rotating stars \citep{Ribas.06, Torres.06, Lopez-Morales.07}. There is also evidence that the discrepancy may be correlated with metallicity \citep{Lopez-Morales.07}. Recently \citet{Morales.07} have shown that for fixed luminosity, active stars tend to have lower temperatures than inactive stars. This result was based on a sample of isolated field stars. 

Our observations show that, at fixed luminosity, rapidly rotating late-K and early-M dwarfs tend to be bluer in $(B-V)$ but redder in $(V-I_{C})$ than slowly rotating dwarfs. Since the bulk of the flux from these stars is emitted in the near infrared, it is reasonable to suppose that the correlation between $(V-I_{C})$ and period more closely represents the correlation between effective temperature and period for these stars than the correlation between $(B-V)$ and the period does. For stars with $8.5 < M_{V} < 9.5$ we find that the slope between $(V-I_{C})$ and the period is $\sim -0.005~{\rm mag}/{\rm day}$. Using the best fit YREC isochrone for M37 (see Paper I), a difference in $(V-I_{C})$ of $0.1~{\rm mag}$ (corresponding to a difference in period of 20 days, which is comparable to the difference between an old field star and a tidally synchronized binary) would result in a $\sim 3\%$ difference in effective temperature at fixed luminosity, or a $\sim 6\%$ difference in radius. This is comparable to, but still slightly less than, the radius discrepancy from eclipsing binaries. Since the flux through the $V$ filter may be slightly enhanced for rapidly rotating stars, it is likely that colors using only near-infrared filters will be more strongly anti-correlated with period than $(V-I_{C})$ is. Deep near-infrared observations of this cluster would confirm or refute this hypothesis. Note that because our present sample of stars are all members of the same cluster, we can rule out age effects as the source of the color discrepancy. This is a conclusion which is not possible to make using samples of field stars. 

\section{Angular Momentum Evolution}

By comparing the distribution of stellar rotation periods between star clusters of different ages we can study the evolution of stellar angular momentum. Both changes in the moment of inertia of a star and changes in its angular momentum contribute to changes in the rotation period. After $\sim 100$ Myr stars have settled onto the main sequence and their moment of inertia changes very little until they evolve onto the sub-giant branch. During this time period the rotation evolution is thought to be dominated by angular momentum loss via a magnetized wind. In this section we compare our observations of M37 with observations of other open clusters to test a simple model of stellar angular momentum evolution.

\subsection{Data for Other Clusters}

Besides M37, there are four clusters older than $\sim 100$ Myr that have significant, publicly available, samples of rotation periods; these are the Pleiades \citep[$100$ Myr;][]{Meynet.93}, NGC 2516 \citep[$140$ Myr;][]{Meynet.93}, M34 \citep[$200$ Myr;][]{Jones.96} and the Hyades \citep[$625$ Myr;][]{Perryman.98}. 

\subsubsection{Pleiades}

We used the WEBDA database\footnote{http://www.univie.ac.at/webda/webda.html} to obtain the rotation periods for 50 Pleiads; the periods are compiled from a number of sources \citep{Stauffer.87a, Prosser.93a, Prosser.93b, Prosser.95, Marilli.97, Krishnamurthi.98, Terndrup.99, Messina.01b, Clarke.04, Scholz.04}. For stars with multiple periods listed we took the average value. We do not include 11 low-mass stars with periods for which optical photometry is unavailable.

We also used WEBDA to obtain the photometry for this cluster. We took the mean photoelectric $BV$ measurements \citep{Johnson.53,Johnson.58,Iriarte.67,Iriarte.74,Mendoza.67,Jones.73,Robinson.74,Landolt.79,Stauffer.80,Stauffer.82b,Stauffer.84,Stauffer.87b,Prosser.91b,Andruk.95,Messina.01b}, Kron $VI_{K}$ measurements \citep{Stauffer.82b, Stauffer.84, Prosser.91b, Stauffer.87b}, Johnson $VI_{J}$ measurements \citep{Mendoza.67, Iriarte.69, Landolt.79} and Johnson-Cousins $VI_{C}$ measurements \citep{Stauffer.98}. Following A07 we converted $VI_{K}$ to $VI_{C}$ using the transformation from \citet{Bessell.87} and $VI_{J}$ to $VI_{C}$ using the transformation given in A07.

To obtain $(B-V)_{0}$, $(V-I_{C})_{0}$ and $M_{V}$ for each star we take $E(B-V) = 0.02$, and $(m-M)_{0} = 5.63$ (A07). We also assume $R_{V} = 3.1$ and $E(V-I_{C})/E(B-V) = 1.37$ (see A07).

\subsubsection{NGC 2516}

The rotation periods for 362 stars in NGC 2516 come from \citet[][hereafter I07]{Irwin.07}. These authors also provide $VI_{C}$ photometry for all of their rotators. We find, however, that when adopting $E(B-V) = 0.125$, $(m-M)_{0} = 8.03$ for this cluster (A07), the $(V-I_{C})_{0}$ vs. $M_V$ lower main sequence falls $\sim 0.1$ mag to the red of the sequence for M37, despite the cluster having a metallicity of $[Fe/H] = -0.07 \pm 0.06$ (A07) or $0.01 \pm 0.07$ \citep{Terndrup.02} compared with $[Fe/H] = 0.045 \pm 0.044$ for M37 (Paper I). When taking the $VI_{C}$ photometry for NGC 2516 from \citet[][hereafter J01]{Jeffries.01} or \citet{Sung.02} the sequence is in good agreement with the M37 sequence. We therefore transform the photometry from I07 to match the J01 photometry via:
\begin{eqnarray}
V_{J01} & = & 1.014V_{I07} - 0.172(V-I_{C})_{I07} + 0.043 \\
I_{C,J01} & = & 0.972I_{C,I07} + 0.040(V-I_{C})_{I07} + 0.314.
\end{eqnarray}
Finally there is $BV$ photometry from J01 for 73 of the I07 rotators.

\subsubsection{M34}

The rotation periods for 105 stars in M34 were taken from \citet[][hereafter I06]{Irwin.06}. We adopt the $VI_{C}$ photometry from this paper as well. The lower main sequence in this case appears to be quite comparable to that for M37 when taking $E(B-V) = 0.07$ \citep{Canterna.79} and $(m-M)_{0} = 8.38$ \citep{Jones.96}. Note that this cluster appears to be slightly more metal rich than M37, with $[Fe/H] = +0.07 \pm 0.04$ \citep{Schuler.03}. There is $BV$ photometry from \citet{Jones.96} for 25 of the I06 rotators. 

\subsubsection{Hyades}

As for the Pleiades we obtained the rotation periods for 25 Hyads from WEBDA; the periods are compiled from three sources \citep{Radick.87,Radick.95,Prosser.95} and we take the average value for stars with multiple measurements. We also take the average photoelectric $BV$ photometry \citep{Johnson.55,Argue.66,Eggen.68,Eggen.74,Mendoza.68,vanAltena.69,Sturch.72,Robinson.74,Upgren.77,Upgren.85,Stauffer.82a,Herbig.86,Weis.82,Weis.88,Andruk.95}, Kron $VI_{K}$ photometry \citep{Upgren.77, Upgren.85, Weis.82, Weis.88, Stauffer.82a}, Johnson $VI_{J}$ photometry \citep{Mendoza.67,Johnson.68,Mendoza.68,Sturch.72,Carney.79}, and Johnson-Cousins $VI_{C}$ photometry \citep{Reid.93}. We transform the $VI_{K}$ and $VI_{J}$ photometry to the $VI_{C}$ system using the same relations that we used for the Pleiades. Finally, we adopt $E(B-V) = 0.003 \pm 0.002$ \citep{Crawford.75, Taylor.80} and an average distance modulus of $(m-M)_{0} = 3.33 \pm 0.01$ \citep{Perryman.98}.

\subsection{Comparison of Period-Age Data with Models}

\begin{figure}[p]
\plotone{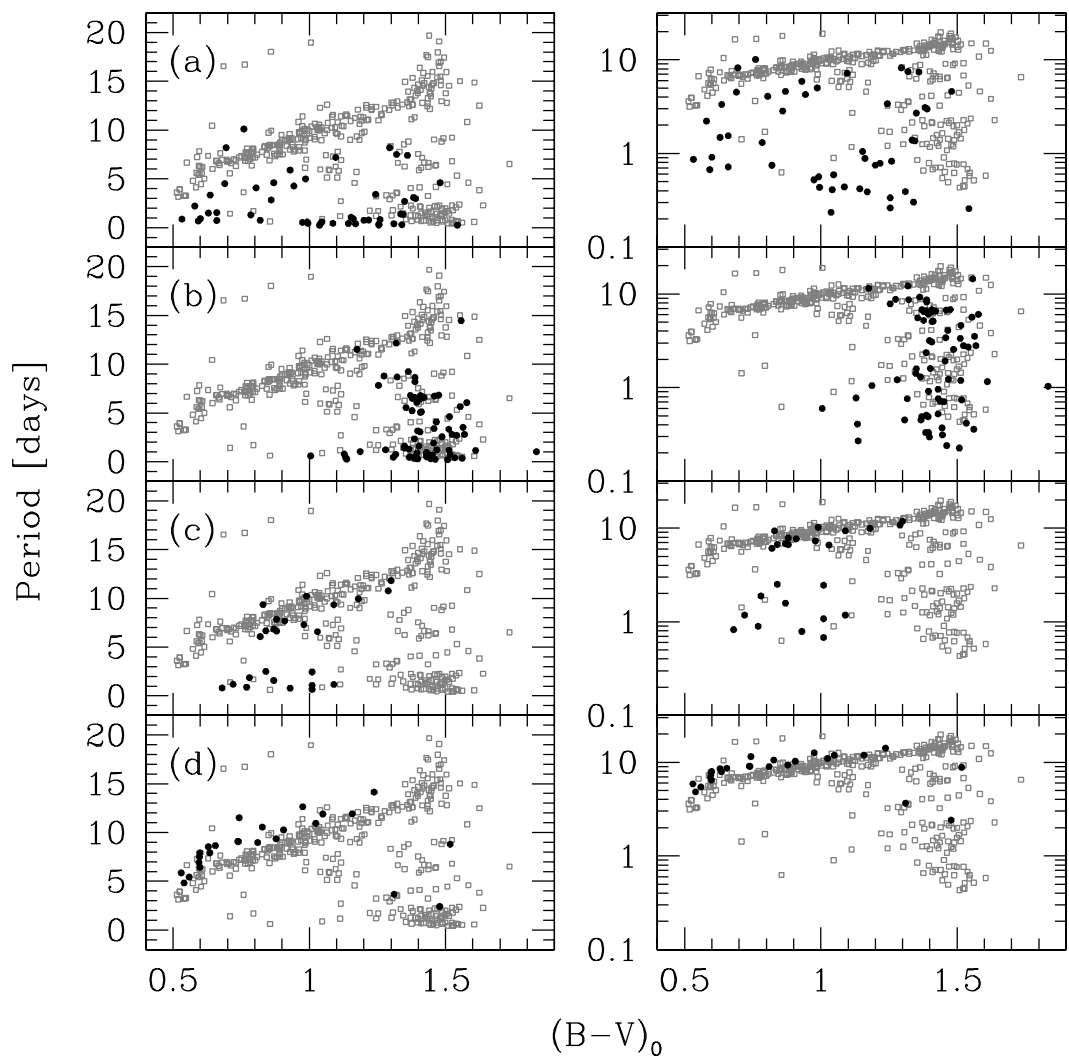}
\caption{Comparison of the $(B-V)_{0}$-period relation between M37 (light open points) and (a) the Pleiades, (b) NGC 2516, (c) M34, and (d) the Hyades. Dark filled points are used for the other clusters. On the left the periods are plotted on a linear scale, on the right they are plotted on a log scale. Note the general evolutionary trend toward longer periods from the youngest cluster (the Pleiades) to the oldest cluster (the Hyades). Note that we use only the clean sample of stars for M37.}
\label{fig:BVPeriod}
\end{figure}

\begin{figure}[p]
\plotone{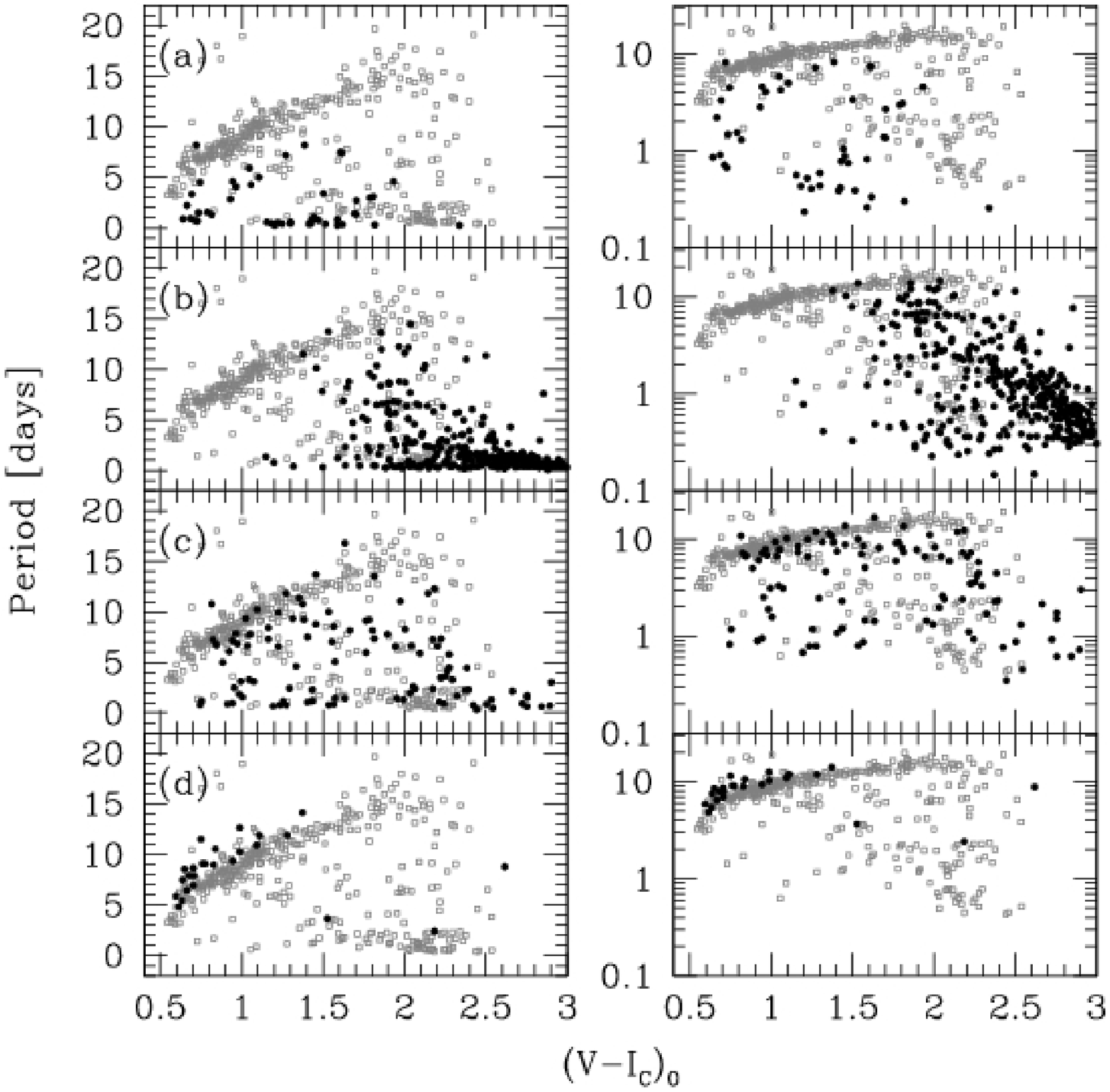}
\caption{Same as figure~\ref{fig:BVPeriod}, here we show the $(V-I_{C})_{0}$-period relation.}
\label{fig:VIPeriod}
\end{figure}

In figure~\ref{fig:BVPeriod} we compare the $(B-V)_{0}$ vs. Period relation for M37 to each of the four clusters discussed above. In figure~\ref{fig:VIPeriod} we show the $(V-I_{C})_{0}$ vs. Period relation. 

The convergence of stellar rotation periods into a sequence for $(B-V)_{0} < 1.3$ is clearly seen in M37 and the Hyades. A sequence may also be present in the Pleiades and M34 (this is more clearly seen in the less complete $BV$ data for M34). The data for NGC 2516 is incomplete for stars hotter than $(B-V)_{0} \sim 1.3$ as the I07 survey was focused on very low mass stars. 

The formation of such a sequence can be explained by an angular momentum loss law that is a steep function of the angular frequency. Typically modelers adopt a modified \citet{Kawaler.88} $N = 1.5$ wind law \citep{Chaboyer.95}, where $N$ is a parameter used in modeling the geometry of the coronal magnetic field of the star and can vary between 0 and 2, with $N = 3/7$ for a dipolar field and $N = 2$ for a purely radial field. The general law has the form:
\begin{equation}
\frac{dJ}{dt} = \left\{ \begin{array}{ll}
	f_{k}K_{w}\omega^{1 + 4N/3}\left(\frac{R}{R_{\odot}}\right)^{2-N}\left(\frac{M}{M_{\odot}}\right)^{-N/3}\left(\frac{\dot{M}}{10^{-14}M_{\odot}{\rm yr}^{-1}}\right)^{1 - 2N/3}, & \omega < \omega_{sat} \\
	f_{k}K_{w}\omega\omega_{crit}^{4N/3}\left(\frac{R}{R_{\odot}}\right)^{2-N}\left(\frac{M}{M_{\odot}}\right)^{-N/3}\left(\frac{\dot{M}}{10^{-14}M_{\odot}{\rm yr}^{-1}}\right)^{1 - 2N/3}, & \omega \geq \omega_{sat} 
\end{array}
\right.
\label{eqn:angmomlossgeneral}
\end{equation}
Here $\omega_{sat}$ is a saturation angular frequency, which is needed to account for the presence of rapid rotators in the Pleiades. The leading constant $f_{k}K_{w}$ determines the overall angular momentum loss rate. \citet{Kawaler.88} gives $f_{k}K_{w} = 2.035 \times 10^{33}(24.93 K_{V}^{-1/2})^{n}K_{B}^{4n/3}$, where $K_{V}$ is the ratio of the wind speed to the escape velocity at a radius of $r_{A}$, the radius out to which the stellar wind co-rotates with the star, and $K_{B}$ is a constant of proportionality between the surface magnetic field strength and the stellar rotation rate. \citet{Chaboyer.95} let $K_{w} = 2.036 \times 10^{33} (1.452 \times 10^{9})^{N}$ in cgs units, for $K_{V} = 1.0$ and $K_{B}$ set to the value obtained from calibration to the solar global magnetic field strength, and introduce $f_{k}$ as a dimensionless parameter to account for our ignorance of $K_{V}$ and $K_{B}$. For $N = 1.5$ this reduces to
\begin{equation}
\frac{dJ}{dt} = \left\{ \begin{array}{ll}
	-K\omega^{3}\left(\frac{R}{R_{\odot}}\right)^{0.5}\left(\frac{M}{M_{\odot}}\right)^{-0.5}, & \omega < \omega_{sat} \\
	-K\omega\omega_{sat}^{2}\left(\frac{R}{R_{\odot}}\right)^{0.5}\left(\frac{M}{M_{\odot}}\right)^{-0.5}, & \omega \geq \omega_{sat}
\end{array}
\right.
\label{eqn:angmomloss}
\end{equation}
where $f_{k}K_{w}$ is combined into a single calibration constant $K$. \citet{Bouvier.97} determined a value of $K = 2.7 \times 10^{47}$ g cm$^{2}$ s by requiring that the law reproduce the rotation frequency of the Sun at $4.5~{\rm Gyr}$ assuming $\omega_{sat} = 14 \omega_{\odot}$ for a $1 M_{\odot}$ star.

For rigid body rotation, the angular frequency obeys the differential equation:
\begin{equation}
\frac{d\omega}{dt} = \frac{1}{I}\frac{dJ}{dt} - \frac{\omega}{I}\frac{dI}{dt}
\label{eqn:domegadt}
\end{equation}
where $I$ is the moment of inertia of the star. Helioseismology suggests that the rigid body rotation approximation is reasonable for the Sun \citep{Gough.90}, though it is uncertain whether this approximation is valid for younger stars. Previous investigations have found that the rigid body approximation reproduces the observed angular velocity distribution of stars older than the Pleiades, while models incorporating internal differential rotation are needed to reproduce the observations of younger clusters \citep[e.g.][]{Sills.00}. In simple models where the core and envelope of the star are assumed to rotate as distinct rigid bodies, $I$ and $\omega$ in equation~\ref{eqn:domegadt} are replaced with $I_{conv}$ and $\omega_{conv}$ because the magnetic wind is tied to the convective envelope. Additional terms are then required to allow for coupling between the core and the envelope (see I07). 

%Solar mass stars older than $\sim 100~{\rm Myr}$ have settled onto the main sequence so the moment of inertia for these stars does not change substantially. For example, at $\sim 200~{\rm Myr}$, $dI/dt \sim 10^{35}$ g cm$^{2}$ s$^{-1}$ for a $1 M_{\odot}$ star (based on evolutionary tracks computed with the YREC isochrones, Terndrup et al. 2007, in preparation), while for $\omega \sim 10^{-5}$ s$^{-1}$, $\omega^{2}K \sim 3 \times 10^{37}$ g cm$^{2}$ s$^{-1}$ so that the first term in equation~\ref{eqn:domegadt} dominates. In that case, integrating equation~\ref{eqn:domegadt} from time $t_{0}$ with angular frequency $\omega_{0} > \omega_{sat}$ to time $t$ yields:
%\begin{equation}
%\omega(t) = \left\{ \begin{array}{ll}
%	\omega_{0} e^{-(t - t_{0})/\tau}, & t \leq t_{sat} \\
%	\frac{\omega_{sat}}{\sqrt{1 + 2(t - t_{sat})/\tau}}, & t > t_{sat}
%\end{array}
%\right.
%\label{eqn:omegaoft}
%\end{equation}
%where $\tau = \frac{I}{K\omega_{sat}^2}(R/R_{\odot})^{-0.5}(M/M_{\odot})^{0.5}$ is the spin-down time-scale for stars in the saturated wind regime and $t_{sat} = t_{0} - \tau\ln(\omega_{sat}/\omega_{0})$ is the time at which the star transitions to the non-saturated spin-down. For lower mass stars the second term in equation~\ref{eqn:domegadt} may be significant even at ages of a few hundred Myr. In this case 

\begin{figure}[p]
\epsscale{0.7}\plotone{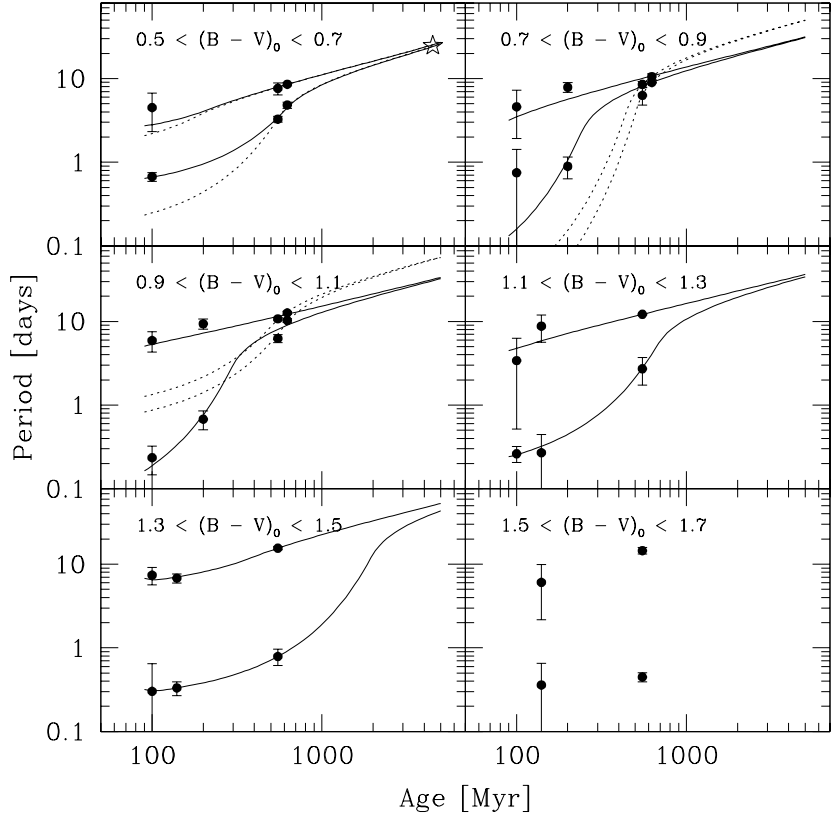}
\caption{The rotation period is plotted as a function of age for stars in the clusters presented in figure~\ref{fig:BVPeriod}. The period-age relation is shown for different $(B-V)_{0}$ color bins. We include the Sun, plotted as an open star, in the $0.5 < (B-V)_{0} < 0.7$ bin, and assume that this is equal to the 10th and 90th percentile of 4.5 Gyr solar-mass stars with an uncertainty of $\sim 1.0~{\rm days}$. The solid points show the 10th and 90th percentile rotation periods for stars in each cluster and color bin. We only show clusters that have at least four points in a given bin. The $1-\sigma$ error bars are estimated by conducting 1000 bootstrap simulations for each cluster. The solid lines show a model fit (equation~\ref{eqn:domegadt}) to the 10th and 90th percentile periods in each color bin for the Sun and the clusters that have at least 4 points. The dotted lines show a fit that is restricted to M37, the Hyades and the Sun (for the bluest color bin) only. While the model can reproduce the observed spin-down and convergence of rotation periods for stars with $0.5 < (B-V)_{0} < 0.7$ or $1.1 < (B-V)_{0} < 1.5$, there are noticeable discrepancies for stars with $0.7 < (B-V)_{0} < 1.1$. We do not attempt to fit the model to color bins with fewer than 3 clusters.}
\label{fig:periodvsageBV}
\end{figure}

\begin{figure}[p]
\epsscale{0.9}\plotone{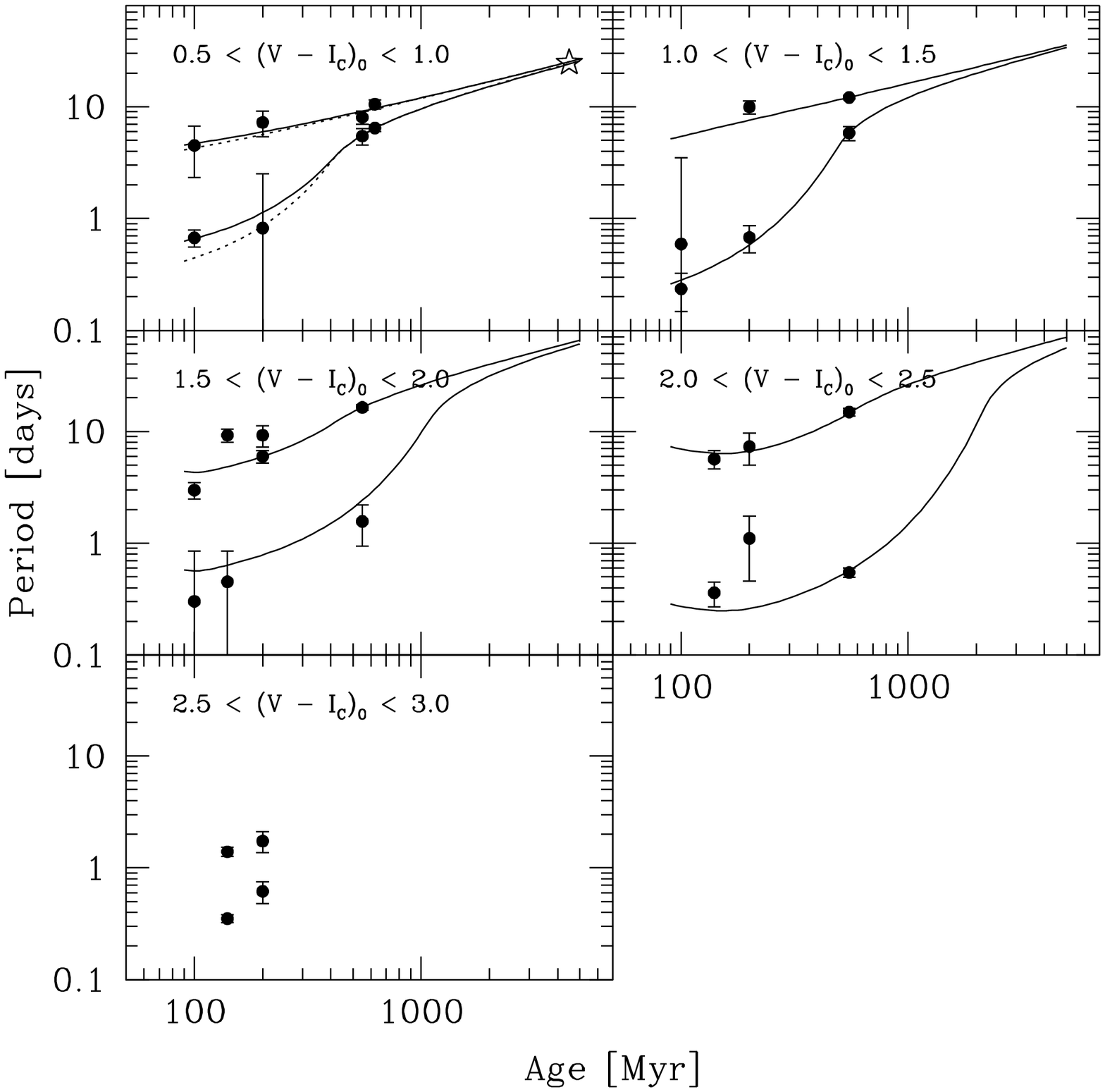}
\caption{Same as figure~\ref{fig:periodvsageBV}, this time for $(V-I_{C})_{0}$. In this case the model provides a good fit to the stars with $0.5 < (V - I_{C})_{0} < 1.0$ and a reasonable fit to the stars with $2.0 < (V-I_{C})_{0} < 2.5$, but fails to fit the stars with $1.0 < (V-I_{C})_{0} < 2.0$ with greater than 95\% confidence. Note that for $1.5 < (V-I_{C})_{0} < 2.5$ the fastest rotators in M34 have periods that are signficantly longer than the fastest rotators in M37. This is particularly apparent in the $1.5 < (V-I_{C})_{0} < 2.0$ bin. The models cannot reproduce the lack of rapid rotators in M34 in these color bins. The models also under-predict the rotation periods of the slowest rotators in M34 with $1.0 < (V-I_{C})_{0} < 2.0$ and the slowest rotators in NGC 2516 with $1.5 < (V-I_{C})_{0} < 2.0$ and over-predict the rotation periods of the slowest rotators in the Pleiades with $1.0 < (V-I_{C})_{0} < 2.0$. We do not attempt to fit the model to color bins with fewer than 3 clusters.}
\label{fig:periodvsageVI}
\end{figure}

In figures~\ref{fig:periodvsageBV}~and~\ref{fig:periodvsageVI} we plot the rotation period as a function of age for stars in the clusters presented in figures~\ref{fig:BVPeriod}~and~\ref{fig:VIPeriod}. Following I07, we show the 10th, and 90th percentile rotation periods for each cluster within color bins. We conduct 1000 bootstrap simulations for each cluster to estimate the $1-\sigma$ uncertainties on these percentiles. For clusters with less than 10 points in a bin, we take the minimum and maximum observed periods in the bin as estimates of the 10th and 90th percentiles. We do not show clusters with less than 4 points in a bin. We include the Sun in the bluest color bins. Note that while the period of the Sun ($P_{\odot} = 24.79~{\rm days}$) is very well determined, we do expect stars at the age of the Sun to exhibit a range in rotation periods. Based on the models described above, we estimate this range to be $\sim 1.0~{\rm day}$ at $4.5~{\rm Gyr}$. We therefore adopt this as the uncertainty on the rotation period of a Sun-like star.

For each color bin we fit a model given eq.~\ref{eqn:domegadt} to the 10th and 90th percentile periods letting $\omega_{0,10}$, $\omega_{0,90}$, $K$, and $\omega_{sat}$ be the free parameters. Here $\omega_{0,10}$ and $\omega_{0,90}$ are the $\omega_{0}$ values at $t_{0} = 100~{\rm Myr}$ for the 10th and 90th percentiles respectively. Note that in solving eq.~\ref{eqn:domegadt} we use evolutionary tracks computed with the YREC isochrones (Terndrup et al. 2008, in preparation) to determine $I$ and $R$ as functions of $M$ and $t$. As seen in figure~\ref{fig:periodvsageBV}, the model can reproduce the observed spin-down and convergence of rotation periods for stars with $0.5 < (B-V)_{0} < 0.7$ or $1.1 < (B-V)_{0} < 1.5$, however we find that for $0.7 < (B-V)_{0} < 1.1$ ($0.76 M_{\odot} < M < 0.99 M_{\odot}$) the model fails to fit the Pleiades, M34, M37 and the Hyades simultaneously with greater than 95\% confidence. In this color-range, the models predict a greater degree of convergence in the rotation periods at the age of M37 than is observed. The models also under-predict the periods of the slowest rotators in M34. When using $(V-I_{C})_{0}$, the models for stars with $1.0 < (V-I_{C})_{0} < 2.5$ ($0.56 M_{\odot} < M < 0.82 M_{\odot}$) fit the M37 observations, but fail to fit the younger clusters. The difference between the $(V-I_{C})_{0}$ and $(B-V)_{0}$ data is that the $(V-I_{C})_{0}$ data is more complete for M34 and NGC 2516 than the $(B-V)_{0}$ data, but is less complete for the Hyades.

Table~\ref{tab:rotparams} gives the parameters for the models displayed in figures~\ref{fig:periodvsageBV}~and~\ref{fig:periodvsageVI}. To simplify the comparison with the data presented in this paper we list periods rather than angular frequencies so that $P_{sat}$ corresponds to $\omega_{sat}$, etc. A few points bear mentioning. The first is that the saturation period appears to increase with decreasing stellar mass, as found by other authors (e.g. I07). The second is that while there is no clear trend between $K$ and stellar mass, we find that to reproduce M37, the Hyades and the Sun $K$ must be a factor of $\sim 1.2$ larger than what was found by \citet{Bouvier.97}.

Throughout this subsection we have implicitely assumed that the samples of stars with rotation period measurements are unbiased in period. Given the relation between period and amplitude (figure~\ref{fig:AmpPerBV}), however, it is likely that the samples are biased towards short-period rotators (though as we saw in \S~\ref{sec:completeness} any bias in period is relatively minor for the M37 data). Since we cannot tell how this bias may differ from sample to sample, it is unclear what affect this has on the above conclusions. Given the fairly sharp upper limit on the period as a function of mass for M37 and the Hyades, we do not expect the estimates of the 90th percentile period measurements for M37 and the Hyades to be substantially different from the actual values for $(B-V)_{0} < 1.1$. We also note that we have not included uncertainties in the fundamental cluster parameters in this analysis. Including these uncertainties will reduce the signficance of the discrepancy between the models and the observed period distributions.

\section{Discussion}

We have measured rotation periods for 575 candidate members of the open cluster M37. This is the largest sample of rotation periods for a cluster older than $100~{\rm Myr}$ and only the second cluster older than $500~{\rm Myr}$ with a large sample of rotation periods (the other cluster being the Hyades with 25 stars that have measured periods). As mentioned in the introduction, this is the second set of rotation periods published for this cluster. \citet{Messina.08} have recently published periods for 122 cluster members. Using this data we have investigated the Rossby number-amplitude and period-color distributions.

We find that for stars with $(B-V)_{0} < 1.36$ the amplitude and Rossby number are anti-correlated, and are related via equation~\ref{eqn:arrofit}. Extrapolating this relation to higher Rossby number values, we expect $A_{r}$ to drop below $1~{\rm mmag}$ at $R_{O} \sim 1.1$. For a Skumanich spin-down, we expect stars to reach this Rossby number at $\sim 2~{\rm Gyr}$. We note that stellar activity will be a non-negligible effect to consider when conducting surveys for transiting planets with amplitudes of $\sim 1~{\rm mmag}$ (e.g. Neptune-sized planets orbiting sun-sized stars) that orbit stars younger than $\sim 2~{\rm Gyr}$. However, since the rotation and transit time-scales typically differ by $\sim 2$ orders of magnitude, it should be possible to find these planets nonetheless. We also find that for stars with $R_{O} < 0.2$, $(B-V)_{0} < 1.36$, the logarithmic distribution of $A_{r}$ appears to be peaked above $0.03~{\rm mag}$ which suggests that the distribution of spot filling-factors is peaked for these rapidly rotating stars.

We have also investigated the effect of rotation on the shape of the main sequence on CMDs. We find that at fixed $V$ magnitude rapid rotators tend to lie blueward of slow rotators on a $B-V$ CMD and redward of slow rotators on a $V-I_{C}$ CMD. This effect, seen previously for early K-dwarfs in the Pleiades and NGC 2516, extends down to early M-dwarf stars ($M_{V} < 10.5$). We note that the relation between $V-I_{C}$ and rotation period is consistent with observations by \citet{Morales.07} who found that at fixed luminosity more active M-dwarfs tend to have lower effective temperatures. Our observations quantitatively and qualitatively support the hypothesis that the well-known discrepancy between the observed and predicted radii and effective temperatures of M-dwarfs is due to stellar activity/rotation and is not an age effect.

We have also investigated the rotation period-color distribution for this cluster. We find that, like the Hyades, the rotation periods of stars in M37 with $0.4 < (B-V)_{0} < 1.0$ follow a tight sequence. At the blue end of this color range stars have rotation periods of $\sim 3~{\rm days}$ while at the red end stars have rotation period of $\sim 10~{\rm days}$. Redward of $(B-V)_{0} \sim 1.0$ the maximum rotation period as a function of color continues to follow this sequence, however there is also a broad tail of stars with shorter rotation periods.  

We have identified a group of 4 stars with $(B-V)_{0} < 1.2$ and $P > 15~{\rm days}$ that fall well above the main period-color sequence. The fact that at least one of these stars, V223, is a likely cluster member warrants discussion. If the measured period corresponds to the rotation period, this star would pose a significant challenge to the theory of stellar angular momentum evolution. While the periods of some of the stars may be shorter than the measured values, the light curve of V223 appears to be well-modeled by a sinusoid with a period of $18.0 \pm 2.9~{\rm days}$ (fig.~\ref{fig:LongPeriodRotsLCS}), the star also has a spectroscopic temperature consistent with its photometric colors assuming that it is a cluster member and $v\sin i = 4.6 \pm 4.6~{\rm km/s}$ that is consistent with a rotation period of $18.0~{\rm days}$. Note that the $r$-band amplitude of the star, $0.024~{\rm mag}$, is an order of magnitude higher than the characteristic amplitude extrapolated to a Rossby number of $0.82$.  

Standard rotation evolution models predict a strong convergence of stellar surface rotation rates for a wide range of initial rotation rates \citep[e.g.][]{Sills.00}. There is no plausible initial rotation rate that would yield a period of $18~{\rm days}$ for V223, which is more than twice as long as the periods of similar stars on the main period-color sequence, at $\sim 500~{\rm Myr}$. To explain such a long rotation period the surface of the star would have to spin-down very rapidly relative to the other stars in the cluster. This may indicate that the internal angular momentum transport varies from star to star as predicted by some theoretical models involving magnetic angular momentum transport \citep[e.g.][]{Charbonneau.93}. Alternatively the slow rotation of these stars might be explained by a threshold effect. For example, stars rotating faster than some threshold may have sufficient mixing to erase gravitational settling, while slower rotators cannot prevent a $\mu$ gradient from being established, and as a result experience core-envelope decoupling. The latter explanation would predict a bimodal distribution of rotation periods, while the former class of models might produce a continuous spectrum.

It is also possible that the variation is due to the evolution of permanently visible spots (e.g. on the pole of an inclined star), rather than the rotational modulation of spots. In that case the measured period corresponds to a spot evolution time-scale rather than to the rotation period.

Finally, we have combined our observations of M37 with previous
observations of the Pleiades, NGC 2516, M34, the Hyades, and the Sun
to perform a test of a simple theory of stellar angular momentum
evolution which assumes rigid body rotation and an $N = 1.5$
wind-model including saturation for short periods. We find that the
model provides a good fit to the data for stars with $0.5 < (B-V)_{0}
< 0.7$ ($0.99 M_{\odot} < M < 1.21 M_{\odot}$) or $1.1 < (B-V)_{0} <
1.5$ ($0.53 M_{\odot} < M < 0.76 M_{\odot}$), but for stars with $0.7
< (B-V)_{0} < 1.1$ ($0.76 M_{\odot} < M < 0.99 M_{\odot}$) the model
fails to fit the Pleiades, M34, M37 and the Hyades simultaneously at
the $2-4\sigma$ level. In this color-range, the best-fit models predict a
greater degree of convergence in the rotation periods at the age of
M37 than is observed. The models also under-predict the periods of the
slowest rotators in M34. When using $(V-I_{C})_{0}$, the models for
stars with $1.0 < (V-I_{C})_{0} < 2.0$ ($0.56 M_{\odot} < M < 0.82
M_{\odot}$) fit the M37 observations, but fail to fit the younger
clusters. Taking the parameters from the best fit models at face
value, we find that the saturation period increases with decreasing
stellar mass, which is consistent with the results from previous
studies (see I07).

Comparing our results to the survey of M37 by
\citet{Messina.08}, we note that both groups have found that there
does not appear to be significant spin-down between M34 and M37 for
stars with $(B-V)_{0} \la 1.0$, while there is significant spin-down
between M37 and the Hyades. While the rotation periods and photometry
are independent for each group, both groups adopt the same fundamental
parameters for the clusters. The results from the two surveys are thus
not independent against errors in these parameters. Our survey goes
more than 2 magnitudes deeper than \citet{Messina.08} (the faintest
star with a measured period in the \citet{Messina.08} survey has $V
\sim 20$, while the faintest star in our survey has $V \sim 22.8$), as
a result we are able to study the rotation evolution for late K and
early M dwarfs as well as late F, G and early K dwarfs (note the
\citet{Messina.08} survey has measured periods for early F stars that
are saturated in our survey). We find that for later spectral-type
stars there is a noticeable spin-down between the slowest rotators in
M34 and M37.

The survey presented in this paper compliments previous surveys of
younger clusters. The recent results from the Monitor project, in
particular, have provided a wealth of data for testing the rotation
evolution of low-mass stars at ages between $5$ and $200~{\rm Myr}$
\citep{Irwin.06,Irwin.07,Irwin.08a,Irwin.08b}. Studying stellar
evolution at these young ages yields insight into processes such as
disk regularization, which may be important for stellar and planetary
formation theory, and internal differential rotation. Other processes,
such as non-saturated angular momentum wind-loss are more significant
for the older cluster studied in this paper.

The discrepancies between this simple theory of stellar angular
momentum evolution and the observed rotation periods of subsolar mass
stars in clusters older than $\sim 100~{\rm Myr}$ may suggest that one
or more of the assumptions in the theory is wrong. In particular, it
may be necessary to relax the assumption that these stars rotate as
rigid bodies, or to revise the assumed wind model. A detailed test of
more complicated models is beyond the scope of this paper. Additional
observations of the rotation periods of stars in clusters with ages
$\ga 100~{\rm Myr}$ and a thorough treatment of all sources of
observational errors are needed to map the period-age evolution in
detail and understand the physical mechanisms behind angular momentum
evolution.

\acknowledgements 
We are grateful to C.~Alcock for providing partial support for this
project through his NSF grant (AST-0501681). Funding for M.~Holman
came from NASA Origins grant NNG06GH69G. We would like to thank
G.~F\"{u}r\'{e}sz and A.~Szentgyorgyi for help in preparing the
Hectochelle observations, S.~Barnes, S.~Saar, N.~Brickhouse and
S.~Baliunas for helpful discussions, and the staff of the MMT, without
whom this work would not have been possible. We are grateful to the
anonymous referee for providing a thoughtful critique which improved
the quality of this paper. We would also like to thank the MMT TAC for
awarding us a significant amount of telescope time for this
project. This research has made use of the WEBDA database, operated at
the Institute for Astronomy of the University of Vienna; it has also
made use of the SIMBAD database, operated at CDS, Strasbourg, France.

\clearpage

\begin{deluxetable}{lrrrrrrrrrrrr}
\tabletypesize{\scriptsize}
\rotate
\tablewidth{0pc}
\tablecaption{Candidate Cluster Members with Measured Rotation Periods: Coordinates and Periods}
\tablehead{
\colhead{ID} & 
\colhead{RA} & 
\colhead{DEC} & 
\colhead{Chip} & 
\colhead{Period\tablenotemark{a}} & 
\colhead{$\sigma_{P}$} & 
\colhead{$P_{N=1}$\tablenotemark{b}} &
\colhead{$\sigma_{P_{N=1}}$} &
\colhead{$P_{N=2}$\tablenotemark{b}} &
\colhead{$\sigma_{P_{N=2}}$} &
\colhead{$P_{N=3}$\tablenotemark{b}} &
\colhead{$\sigma_{P_{N=3}}$} &
\colhead{Clean\tablenotemark{c}} \\
 & \multicolumn{1}{c}{(J2000)} & \multicolumn{1}{c}{(J2000)} & & \multicolumn{1}{c}{(days)} & \multicolumn{1}{c}{(days)} & \multicolumn{1}{c}{(days)} & \multicolumn{1}{c}{(days)} & \multicolumn{1}{c}{(days)} & \multicolumn{1}{c}{(days)} & \multicolumn{1}{c}{(days)} & \multicolumn{1}{c}{(days)} &}
\startdata
    25 & 05:51:17.45 & +32:40:08.7 & 29 &  7.96596671 &  0.02481300 &  7.93346085 &  0.01864600 & 15.94418247 &  0.04626800 & 15.78239472 &  1.78339100 & 0 \\
    26 & 05:51:18.34 & +32:40:31.9 & 29 &  8.21743495 &  0.01027100 &  8.38240366 &  0.01186800 &  8.35987638 &  0.02384200 &  8.35987638 &  0.02914300 & 1 \\
    43 & 05:51:20.55 & +32:32:01.6 & 32 &  6.32258130 &  0.01705700 &  6.38618354 &  0.01925000 & 12.74358869 &  0.04080600 & 12.90215678 &  0.02950900 & 0 \\
    47 & 05:51:21.37 & +32:35:39.1 & 31 &  1.69921452 &  0.00069700 &  1.70573845 &  0.00064600 &  1.70107340 &  0.00045400 &  1.70107340 &  0.00045400 & 1 \\
    57 & 05:51:22.06 & +32:31:02.6 & 33 &  7.62148720 &  0.03865000 &  7.73616016 &  0.02092800 &  7.77483129 &  0.02163400 &  7.75544752 &  0.01871000 & 1 \\
    58 & 05:51:22.14 & +32:29:42.4 & 33 & 15.58862440 &  0.25834300 & 17.36856505 &  1.97524300 & 16.71526061 &  3.35777700 & 16.19318063 &  2.72763000 & 1 \\
    61 & 05:51:22.34 & +32:37:54.3 & 30 &  8.45071991 &  0.01507800 &  4.17488928 &  0.00342200 &  8.34977855 &  0.00851300 & 12.47444192 &  0.02091000 & 0 \\
    71 & 05:51:24.23 & +32:38:01.3 & 30 &  8.73549565 &  0.02335200 &  8.66251740 &  0.02124000 &  8.47374013 &  0.01470200 &  8.71103332 &  0.04233500 & 1 \\
\enddata
\tablenotetext{a}{The period from Paper II.}
\tablenotetext{b}{The period from the multi-harmonic AoV algorithm, with the specified value of $N$.}
\tablenotetext{c}{A flag that is either 1 or 0. Stars with a value of 1 were included in the clean sample of stars in \S 5.}
\label{tab:M37_rotation1}
\end{deluxetable}

\begin{deluxetable}{lrrrrrrrrrrrrrr}
\tabletypesize{\scriptsize}
\rotate
\tablewidth{0pc}
\tablecaption{Candidate Cluster Members with Measured Rotation Periods: Magnitudes}
\tablehead{
\colhead{ID} & 
\colhead{$<\!r\!>$\tablenotemark{a}} & 
\colhead{$RMS_{r}$\tablenotemark{b}} &
\colhead{$g$\tablenotemark{c}} &
\colhead{$r$\tablenotemark{c}} &
\colhead{$i$\tablenotemark{c}} &
\colhead{$B$\tablenotemark{d}} &
\colhead{$V$\tablenotemark{d}} &
\colhead{$I$\tablenotemark{e}} &
\colhead{$A_{r,N=1}$\tablenotemark{f}} &
\colhead{min $A_{r,N=1}$\tablenotemark{g}} &
\colhead{$A_{r,N=2}$\tablenotemark{f}} &
\colhead{min $A_{r,N=2}$\tablenotemark{g}} &
\colhead{$A_{r,N=3}$\tablenotemark{f}} &
\colhead{min $A_{r,N=3}$\tablenotemark{g}} \\
 & \multicolumn{1}{c}{(mag)} & \multicolumn{1}{c}{(mag)} & \multicolumn{1}{c}{(mag)} & \multicolumn{1}{c}{(mag)} & \multicolumn{1}{c}{(mag)} & \multicolumn{1}{c}{(mag)} & \multicolumn{1}{c}{(mag)} & \multicolumn{1}{c}{(mag)} & \multicolumn{1}{c}{(mag)} & \multicolumn{1}{c}{(mag)} & \multicolumn{1}{c}{(mag)} & \multicolumn{1}{c}{(mag)}}
\startdata
    25 & 16.743 &  0.016 & 17.548 & 16.732 & 16.401 & 18.064 & 17.076 & 15.966 &  0.027 &  0.020 &  0.031 &  0.024 &  0.032 &  0.024 \\
    26 & 17.341 &  0.005 & 18.365 & 17.344 & 16.881 & 18.879 & 17.735 & 16.438 &  0.013 &  0.006 &  0.013 &  0.006 &  0.014 &  0.006 \\
    43 & 16.423 &  0.007 & 17.198 & 16.421 & 16.074 & 17.707 & 16.763 & 15.637 &  0.011 &  0.008 &  0.013 &  0.010 &  0.013 &  0.009 \\
    47 & 18.930 &  0.021 & 20.303 & 18.923 & 18.131 & 21.125 & 19.645 & 17.673 &  0.051 &  0.015 &  0.059 &  0.012 &  0.059 &  0.012 \\
    57 & 16.972 &  0.005 & 17.823 & 16.974 & 16.576 & 18.347 & 17.325 & 16.137 &  0.009 &  0.005 &  0.010 &  0.005 &  0.011 &  0.006 \\
    58 & 16.743 &  0.006 & 17.566 & 16.742 & 16.311 & 18.089 & 17.099 & 15.871 &  0.012 &  0.005 &  0.014 &  0.005 &  0.015 &  0.005 \\
    61 & 19.725 &  0.019 & 21.366 & 19.721 & 18.584 & 21.958 & 20.305 & 18.110 &  0.045 &  0.014 &  0.046 &  0.014 &  0.049 &  0.015 \\
    71 & 17.444 &  0.006 & 18.439 & 17.445 & 16.991 & 19.042 & 17.877 & 16.550 &  0.012 &  0.006 &  0.016 &  0.007 &  0.016 &  0.006 \\
\enddata
\tablenotetext{a}{The flux averaged $r$ magnitude of the light curve.}
\tablenotetext{b}{The root-mean-square of the light curve.}
\tablenotetext{c}{The magnitude of the source in the photometric catalog (Paper I)}
\tablenotetext{d}{Value from \citet{Kalirai.01}}
\tablenotetext{e}{Value transformed from $r$ and $i$}
\tablenotetext{f}{Peak-to-peak amplitude in $r$-band for the multi-harmonic AoV model with specified $N$.}
\tablenotetext{g}{Minimum peak-to-peak amplitude in $r$-band that the light curve could have had and still have been detected as a variable. Entries with no value correspond to light curves that would be selected as variable even when $A_{r,N} = 0.0$. The model does not adequately describe the light curve in these cases.}
\label{tab:M37_rotation2}
\end{deluxetable}

\begin{deluxetable}{lrrrrrr}
\tabletypesize{\scriptsize}
\rotate
\tablewidth{0pc}
\tablecaption{Candidate Cluster Members with Measured Rotation Periods: Spectroscopic Parameters}
\tablehead{
\colhead{ID} & 
\colhead{$T_{eff}$} &
\colhead{$\sigma T_{eff}$} &
\colhead{$v \sin i$} &
\colhead{$\sigma v \sin i$\tablenotemark{a}} &
\colhead{$RV$} &
\colhead{$\sigma RV$} \\
 & \multicolumn{1}{c}{(K)} & \multicolumn{1}{c}{(K)} & \multicolumn{1}{c}{(km/s)} & \multicolumn{1}{c}{(km/s)} & \multicolumn{1}{c}{(km/s)} & \multicolumn{1}{c}{(km/s)}}
\startdata
    25 & 5384 &  123 &   7.53 &   1.20 &   8.65 &  0.190 \\
    26 & 5057 &  122 &  11.94 &   1.81 &   9.38 &  0.322 \\
    71 & 4991 &  133 &   8.06 &   1.54 &   8.68 &  0.108 \\
    99 & 4785 &  378 &  10.94 &  19.02 &   7.60 &  0.758 \\
   103 & 5216 &  127 &   3.65 &   1.78 &   8.90 &  0.237 \\
   104 & 4773 & 1506 &  12.04 & 418.07 &  48.85 & 15.984 \\
   174 & 4741 &  584 &   8.72 &  35.15 &  -1.76 &  0.805 \\
   216 & 6521 &  287 &  48.36 &   3.19 &   9.02 &  0.242 \\
\enddata
\tablenotetext{a}{In cases where $\sigma v \sin i$ exceeds $v \sin i$, the values for $\sigma v \sin i$ should be taken as an upper limit on $v \sin i$.}
\label{tab:M37_rotation3}
\end{deluxetable}

\begin{deluxetable}{rrrrr}
\tabletypesize{\scriptsize}
\tablewidth{0pc}
\tablecaption{Color Uncertainty as a Function of Absolute Magnitude}
\tablehead{\colhead{$M_{V} (mag)$} & \colhead{$\sigma(B-V)_{0}$ (mag)} & \colhead{$\sigma(V-I_{C})_{0}$ (mag)} & \colhead{$\sigma(g-r)$ (mag)} & \colhead{$\sigma(g-i)$ (mag)}}
\startdata
 4.0 & 0.012 & 0.027 & 0.032 & 0.034 \\
 5.0 & 0.017 & 0.046 & 0.035 & 0.049 \\
 6.0 & 0.016 & 0.036 & 0.034 & 0.046 \\
 7.0 & 0.022 & 0.045 & 0.046 & 0.062 \\
 8.0 & 0.035 & 0.062 & 0.066 & 0.073 \\
 9.0 & 0.039 & 0.107 & 0.069 & 0.094 \\
10.0 & 0.066 & 0.147 & 0.059 & 0.102 \\
11.0 & 0.075 & 0.069 & 0.061 & 0.060 \\
\enddata
\label{tab:coloruncertainties}
\end{deluxetable}

\begin{deluxetable}{llll}
\tabletypesize{\scriptsize}
\tablewidth{0pc}
\tablecaption{Parameters from Fitting equations~\ref{eqn:M37sequence_linear}~and~\ref{eqn:M37sequence_bpl}.}
\tablehead{
\colhead{Parameter} & \colhead{Value} & \colhead{Bootstrap Error} & \colhead{Monte Carlo Error}
}
\startdata
$a_{lin}$ &  10.81 days/mag &    0.35  &    0.14 \\
$b_{lin}$ &  -1.36 days     &    0.35  &    0.12 \\
$a_{bpl}$ &   4.752 days    &    0.031 &    0.021 \\
$b_{bpl}$ & -20.9           &    4.0   &    2.0 \\
\enddata
\label{tab:M37sequencefit}
\end{deluxetable}

\begin{deluxetable}{rrrrrrrr}
\tabletypesize{\scriptsize}
\tablewidth{0pc}
\tablecaption{Spread in Period about the Main Period-Color Band as a Function of Color.}
\tablehead{\colhead{$(B-V)_{0}$} & \colhead{\# Points} & \colhead{$RMS$\tablenotemark{a}} & \colhead{$\chi^{2}$\tablenotemark{b}} & \colhead{significance\tablenotemark{c}} & \colhead{$RMS$\tablenotemark{d}} & \colhead{$\chi^{2}$\tablenotemark{e}} & \colhead{significance\tablenotemark{f}}\\
& & \multicolumn{1}{c}{linear (days)} & \multicolumn{1}{c}{linear} & \multicolumn{1}{c}{linear} & \multicolumn{1}{c}{B.P.L. (days)} & \multicolumn{1}{c}{B.P.L.} & \multicolumn{1}{c}{B.P.L.}}
\startdata
0.60 &  33 & 1.07 & 10.77 & 39.51 &  9.28 & 5.11 & 16.62 \\
0.80 &  51 & 0.54 & 1.69 &  3.46 &  2.58 & 1.65 & 3.25 \\
1.00 &  61 & 0.77 & 2.22 &  6.73 &  4.34 & 2.34 & 7.38 \\
1.20 &  35 & 0.72 & 2.90 &  7.92 &  5.52 & 2.71 & 7.11 \\
1.40 &  57 & 1.36 & 1.57 &  3.03 &  3.03 & 1.80 & 4.24 \\
1.60 &   7 & 0.69 & 2.72 &  3.21 &  2.79 & 1.93 & 1.73 \\
\enddata
\tablenotetext{a}{Standard deviation of the period residuals of stars within 0.1 mag of $(B-V)_{0}$ for a linear fit between period and $(B-V)_{0}$.}
\tablenotetext{b}{$\chi^{2}$ per degree of freedom of the period residuals of stars within 0.1 mag of $(B-V)_{0}$ for a linear fit between period and $(B-V)_{0}$}
\tablenotetext{c}{Significance in $\sigma$ of $\chi^{2}$ for the linear model.}
\tablenotetext{d}{Standard deviation of the period residuals of stars within 0.1 mag of $(B-V)_{0}$ for a broken power-law fit between period and $(B-V)_{0}$.}
\tablenotetext{e}{$\chi^{2}$ per degree of freedom of the period residuals of stars within 0.1 mag of $(B-V)_{0}$ for a broken power-law fit between period and $(B-V)_{0}$}
\tablenotetext{f}{Significance in $\sigma$ of $\chi^{2}$ for the broken power-law model.}
\label{tab:M37sequence}
\end{deluxetable}

\begin{deluxetable}{rrrrrrrrrr}
\tabletypesize{\scriptsize}
\tablewidth{0pc}
\tablecaption{Parameters for models displayed in figures~\ref{fig:periodvsageBV}~and~\ref{fig:periodvsageVI}}
\tablehead{
\colhead{Color Range} & \colhead{Mass Range\tablenotemark{a}} & \colhead{$P_{0,10}$\tablenotemark{b}} & \colhead{$P_{0,90}$\tablenotemark{b}} & \colhead{$P_{sat}$\tablenotemark{b}} & \colhead{$\omega_{sat}/\omega_{\odot}$} & \colhead{$K$\tablenotemark{b}} & \colhead{$\chi^{2}$} & \colhead{D.O.F.\tablenotemark{c}} & \colhead{C.L.\tablenotemark{d}}\\
\multicolumn{1}{c}{(mag)} & \multicolumn{1}{c}{($M_{\odot}$)} & \multicolumn{1}{c}{(days)} & \multicolumn{1}{c}{(days)} & \multicolumn{1}{c}{(days)} & & \multicolumn{1}{c}{($10^{47}$ g cm$^{2}$ s)} & & &}
\startdata
\cutinhead{$(B-V)_{0}$, using data from all available clusters and the Sun\tablenotemark{e}}
$[0.5, 0.7]$ & $[0.99, 1.21]$ &  $0.64 (0.06)$ & $ 2.71 (0.50)$ & $ 4.26 (0.18)$ &  5.82 & $ 3.25 (0.18)$ &  2.39 &  4 & 0.43 \\
$[0.7, 0.9]$ & $[0.86, 0.99]$ &  $0.13 (0.04)$ & $ 3.18 (0.84)$ & $ 2.28 (0.27)$ & 10.89 & $ 3.65 (0.18)$ & 12.21 &  4 & 2.41 \\
$[0.9, 1.1]$ & $[0.76, 0.86]$ &  $0.16 (0.05)$ & $ 5.06 (1.01)$ & $ 2.85 (0.32)$ &  8.70 & $ 3.24 (0.27)$ & 21.75 &  4 & 3.69 \\
$[1.1, 1.3]$ & $[0.68, 0.76]$ &  $0.24 (0.05)$ & $ 4.48 (0.74)$ & $ 5.06 (1.27)$ &  4.90 & $ 3.16 (0.36)$ &  1.19 &  2 & 0.59 \\
$[1.3, 1.5]$ & $[0.53, 0.68]$ &  $0.32 (0.05)$ & $ 6.76 (0.74)$ & $12.15 (1.85)$ &  2.04 & $ 5.35 (1.09)$ &  0.31 &  2 & 0.18 \\
\cutinhead{$(B-V)_{0}$, using data from M37, the Hyades, and the Sun only}
$[0.5, 0.7]$ & $[0.99, 1.21]$ &  $0.23 (0.05)$ & $ 2.07 (0.55)$ & $ 3.41 (0.15)$ &  7.26 & $ 3.29 (0.18)$ &  1.07 &  2 & 0.54 \\
\cutinhead{$(V-I_{C})_{0}$, using data from all available clusters and the Sun}
$[0.5, 1.0]$ & $[0.82, 1.35]$ &  $0.63 (0.11)$ & $ 4.55 (0.99)$ & $ 3.56 (0.34)$ &  6.97 & $ 2.85 (0.16)$ &  2.96 &  6 & 0.24 \\
$[1.0, 1.5]$ & $[0.67, 0.82]$ &  $0.26 (0.06)$ & $ 5.16 (1.24)$ & $ 4.25 (0.48)$ &  5.83 & $ 3.11 (0.44)$ &  6.59 &  2 & 2.08 \\
\enddata
\tablenotetext{a}{The rough range in masses corresponding to the color range of the bin.}
\tablenotetext{b}{Estimates of the $1-\sigma$ uncertainties on the parameters are given in parentheses. These uncertainties are determined by conducting an MCMC simulation.}
\tablenotetext{c}{Number of degrees of freedom in the fit.}
\tablenotetext{d}{The formal confidence level in $\sigma$ at which the null hypothesis that the model fits the data can be rejected.}
\tablenotetext{e}{We exclude clusters with fewer than 4 points in the color bin from the fit. The Sun is only included for the bluest color bin.}
\label{tab:rotparams}
\end{deluxetable}
\end{document}